\documentclass[journal]{IEEEtran}
\usepackage{cite}

\usepackage{amsfonts}

\ifCLASSINFOpdf
   \usepackage[pdftex]{graphicx}
   \graphicspath{{../pdf/}{../jpeg/}}
   \DeclareGraphicsExtensions{.pdf,.jpeg,.png}
\else
   \usepackage[dvips]{graphicx}
   \graphicspath{{../eps/}}
   \DeclareGraphicsExtensions{.eps}
\fi
\usepackage{amsmath}
\usepackage{multirow}
\usepackage{amssymb}
\usepackage{bbding}
\usepackage{pifont}
\usepackage{booktabs} 
\usepackage{tabularx} 
\usepackage{makecell} 
\usepackage{array}
\usepackage{color}

\ifCLASSOPTIONcompsoc
  \usepackage[caption=false,font=normalsize,labelfont=sf,textfont=sf]{subfig}
\else
  \usepackage[caption=false,font=footnotesize]{subfig}
\fi
\usepackage[hyphens]{url}
\usepackage{hyperref}
\hypersetup{hidelinks}

\makeatletter

\newcommand{\Rmnum}[1]{\expandafter\@slowromancap\romannumeral #1@}
\makeatother

\usepackage{graphicx}
\usepackage{amssymb}
\usepackage{amsfonts}
\usepackage{tabularx}
\usepackage{booktabs} 
\usepackage{ifpdf}
\usepackage{cite}
\usepackage{algorithmic}
\usepackage[caption=false,font=footnotesize]{subfig}

\usepackage{multicol}
\usepackage{color}
\usepackage{amsmath}
\usepackage{multirow}
\usepackage{threeparttable}
\begin{document}
%

\title{Phased Array Calibration based on Rotating-Element Harmonic Electric-Field Vector with Time Modulation}
%
%
%

\author{Shiyuan Li, Yuyue Zhou, Chi Zhang, Liang Kong, Kebin Liu, Yihan Xie, Chong He,~\IEEEmembership{Member,~IEEE}
\thanks{This work was supported by the National Natural Science Foundation of
	China 61901263, 62001291. \emph{(Corresponding author: Chong He.)}}
\thanks{The authors are with the Department of Electronic Engineering, Shanghai Jiao Tong University, Shanghai 200240, China (e-mail: hechong@sjtu.edu.cn).}
\thanks{Color versions of one or more of the figures are available online at http:// ieeexplore.ieee.org. (\textbf{NOTE:} Only Used with Printed Publications).}}

\maketitle

\begin{abstract}
Calibration is crucial for ensuring the performance of phased
array since amplitude-phase imbalance between elements results in significant performance degradation. While amplitude-only calibration methods offer advantages when phase measurements are impractical, conventional approaches face two key challenges: they typically require high-resolution phase shifters and remain susceptible to phase errors inherent in these components. To overcome these limitations, we propose a Rotating element Harmonic Electric-field Vector (RHEV) strategy, which enables precise calibration through time modulation principles. The proposed technique functions as follows. Two 1-bit phase shifters are periodically phase-switched at the same frequency, each generating corresponding harmonics. By adjusting the relative delay between their modulation timings, the phase difference between the $+1$st harmonics produced by the two elements can be precisely controlled, utilizing the time-shift property of the Fourier transform. Furthermore, the $+1$st harmonic generated by sequential modulation of individual elements exhibits a linear relationship with the amplitude of the modulated element, enabling amplitude ambiguity resolution. The proposed RHEV-based calibration method generates phase shifts through relative timing delays rather than physical phase shifter adjustments, rendering it less susceptible to phase shift errors. Additionally, since the calibration process exclusively utilizes the $+1$st harmonic, which is produced solely by the modulated unit, the method demonstrates consistent performance regardless of array size. Extensive numerical simulations, practical in-channel and over-the-air (OTA) calibration experiments demonstrate the effectiveness and distinct advantages of the proposed method. 
\end{abstract}


%
\IEEEpeerreviewmaketitle

\section{Introduction}
%
%
%
%
\IEEEPARstart{P}{hased} arrays have been extensively utilized in applications such as autonomous driving \cite{9172393, 9127853}, satellite communication \cite{9711526, 7004863}, and 6G networks \cite{10005812} due to their exceptional beam-steering flexibility. These beamforming capabilities are primarily enabled by digitally controlled phase shifters and amplifiers (or attenuators) integrated into each antenna's radio frequency (RF) chain \cite{9431755}. Through the adjustment of these phase shifters and attenuators, functions such as signal enhancement and anti-interference can be realized \cite{9143417, 1438487}. However, amplitude-phase imbalances inevitably arise across RF chains due to manufacturing tolerances, assembly variations, and environmental fluctuations. Consequently, accurate characterization and compensation of these imbalances are imperative to maintain optimal system performance \cite{10636058}, which is collectively termed phased array calibration.

Existing phased array calibration methods can be primarily classified into several categories: park and probe method \cite{9022967, 9335235}, phase toggling method \cite{789490, 5979144, 7902129, 9521756, 9423585} , mutual coupling method \cite{29378, 859012, 6507334, 5313929, 7323833}, time modulation method \cite{6868196, 10310232}, rotating element electric field vector (REV) method \cite{mano1982method, tanaka1998orbit, 6193154, 7745942, 9445668, 10326459}. Depending on the distance between the calibration source and the antenna under test (AUT), these methods can be grouped into near-field \cite{9022967, 10636058, 768795}, mid-field \cite{8708951} and far-field \cite{9335235, mano1982method, tanaka1998orbit, 6193154, 7745942, 9445668, 10326459, 6868196, 10310232} calibration methods. In terms of the data required for the calibration process, the methods can be classified as complex (amplitude and phase) \cite{9022967, 9335235, 789490, 5979144, 7902129, 6868196, 10310232} or amplitude-only \cite{mano1982method, tanaka1998orbit, 6193154, 7745942, 9445668, 10326459} calibration methods.

The park and probe method \cite{9022967, 9335235} is also known as the near-field scanning probe method. Although this method is intuitive, however, its calibration accuracy heavily depends on the precise relative positioning between the probe and each element. The phase toggling method formulates a linear equation relating the AUT’s phase setting matrix, the complex excitations of the antenna elements, and the received complex signals. The complex excitations of the antenna elements are determined by solving this equation. In \cite{7902129}, the phase settings were designed using a Hadamard matrix, allowing N-element calibration with only a 1-bit phase shifter and one auxiliary antenna (N measurements required). In \cite{9521756, 9423585}, for arrays restricted to beam-scanning mode, the calibration was achieved by deploying multiple symmetric auxiliary antennas. The mutual coupling method calibrates phased arrays by exploiting the natural symmetry between uniformly spaced antenna elements, eliminating the need for external probes. The foundational work in \cite{29378} first proposed calibrating phased arrays by measuring mutual coupling coefficients between elements. In \cite{1256985}, active uniform planar arrays (UPAs) were divided into sub-arrays with passive central calibration units. The method was extended to digital phased arrays in \cite{5313929}. While advantageous by using intrinsic array properties, this method requires perfectly matched antennas with uniform spacing and simultaneous transmit/receive capability, increasing design complexity.  

The time modulation method enables simultaneous calibration of multiple antenna elements in the AUT. In \cite{6868196}, the amplitude-phase response imbalances across four channels were derived by analyzing the harmonic characteristics of the received signal. To relax the dynamic range requirements of the analog-to-digital converter (ADC), \cite{10310232} employed distinct modulation frequencies for each RF channel. While this approach facilitates parallel calibration of array elements, its calibration accuracy is influenced by the precision of phase shifters. Moreover, it necessitates strict synchronous sampling \cite{10040910, 9201395}. Notably, like other techniques, it still relies on high-accuracy phase measurements.

The above phased array calibration methods typically require precise phase measurements, which can be particularly challenging to implement for on-site or millimeter-wave systems. In contrast, amplitude-only approaches such as the rotating element electric field vector (REV) method provide a more practical alternative that only requires standard equipment like spectrum analyzers. The REV method implemented by measuring the power response of the array while incrementally shifting the phase of each element from $0^\circ$ to $360^\circ$. This technique has seen significant developments, including Fourier-based estimation of complex fields derived from power-phase curves \cite{6193154}, a faster calibration approach using just three phase states ($0^\circ$, $90^\circ$, $180^\circ$) \cite{7745942}, algorithmic improvements incorporating variable attenuators to eliminate solution ambiguities \cite{9445668}, and group-wise implementations that enable calibration of large-scale arrays even under low-SNR conditions \cite{10326459}. However, the calibration accuracy of the above amplitude-only methods heavily depends on the phase shifter accuracy, where the error in commonly commercial 6-bit digital phase shifters (such as HMC642ALC5 made by Analog Devices Inc.) reaches up to $11^\circ$ and seriously mitigates the calibration performance. Furthermore, the requirement for at least three distinct phase states makes them unsuitable for systems employing few-bit phase shifters.

To address these limitations, we propose a calibration scheme named RHEV by exploiting the time-modulation strategy, i.e., a \textbf{R}otating element \textbf{H}armonic \textbf{E}lectric-field \textbf{V}ector strategy. The core idea of RHEV involves generating precise phase shifts at the $+1$st harmonic through periodic modulation of 1-bit phase shifters between array elements. This idea is fundamentally inspired by the time-shift property of Fourier transforms, which mathematically establishes that the time-domain delay introduces a linear phase shift in the frequency domain. According to this property, the phase of harmonics in a time-modulated array can be precisely controlled by the delay between elements \cite{10681654}. Building on our previous work \cite{10242122}, which demonstrated an order-of-magnitude reduction in phase shift RMSE compared to conventional digital phase shifters using time modulation techniques. While RHEV shares the amplitude ratio ambiguity characteristic of conventional REV methods, this limitation is effectively resolved by analyzing the $+1$st harmonic component generated through sequential channel modulation. This harmonic analysis enables unambiguous determination of amplitude ratios, thereby eliminating the fundamental constraint of traditional approaches. 

The main contributions are summarized as follows:

(1) We propose an innovative phased array OTA calibration method using only harmonic amplitudes and establish a detailed signal model. The calibration accuracy is robust against phase shift imperfections, as the method introduces phase variation through precisely controlled time delays rather than relying on analog phase shifters. In addition, this method achieves calibration by utilizing only the power of the $+1$st harmonic, which is generated solely by the modulated elements, while the remaining unmodulated elements do not contribute to the $+1$st harmonic power. Therefore, the calibration performance is not much affected by the array size.

(2) The proposed calibration method resolves amplitude ratio ambiguity by exploiting a key characteristic: when individual elements are modulated sequentially, the resulting $+1$st harmonic amplitude maintains direct proportionality with the modulated element's amplitude response. This proportional relationship enables unambiguous amplitude determination during the calibration process.

(3) We designed, fabricated, and experimentally validated a prototype system capable of high-precision +1st harmonic phase control through relative time delay adjustment, accurate in-channel amplitude-phase imbalance measurements, and successful OTA array calibration. These combined achievements demonstrate the effectiveness of the proposed methodology.

The reminder of this paper is organized as follows. Section II presents the mathematical foundation and theoretical analysis of the proposed method. Section III evaluates its performance through numerical simulations. Section IV validates the method with experimental results. Finally, Section V concludes the paper and discusses future research directions.

\begin{figure*}[!t]
	\centering
	\includegraphics[width=12cm]{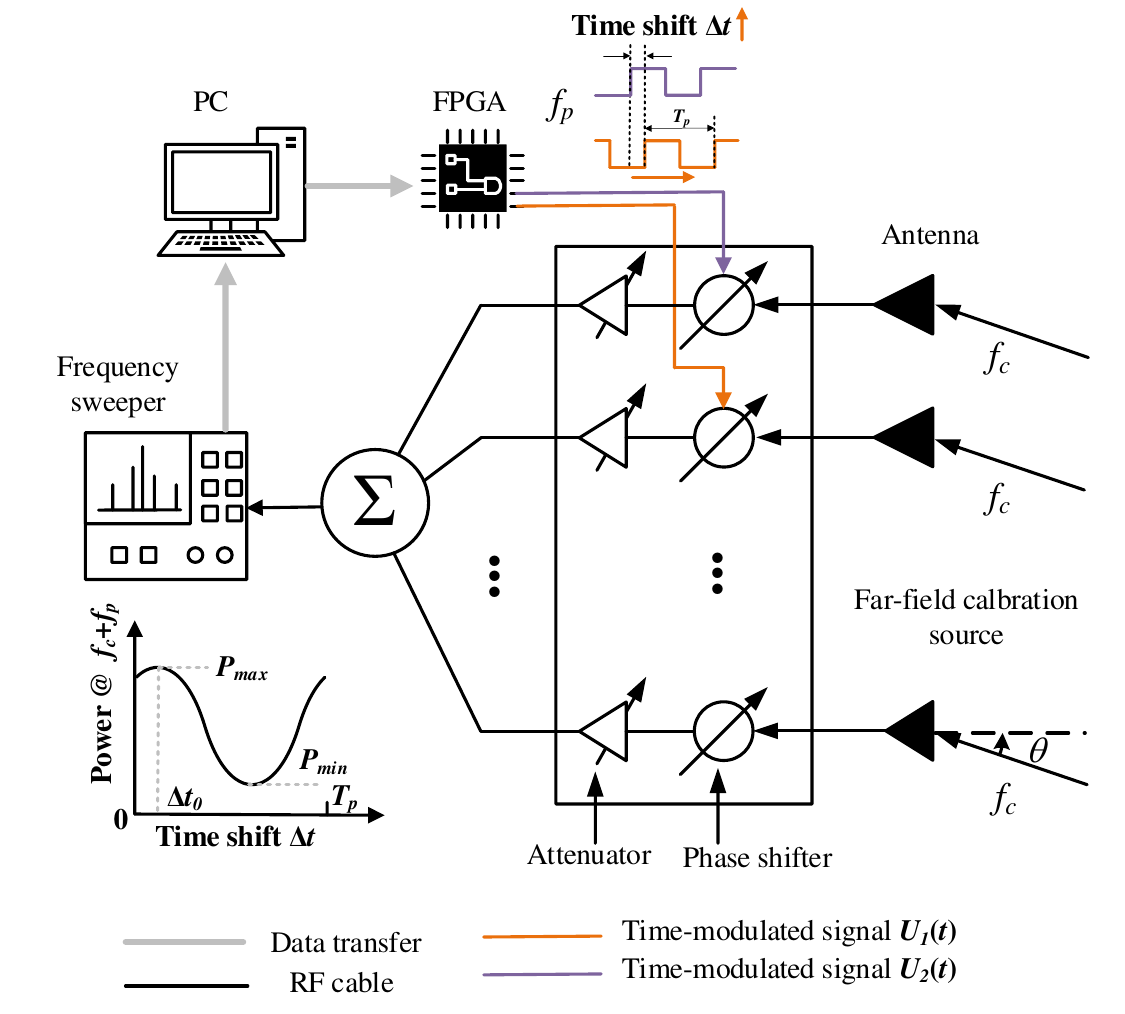}
	\caption{Flowchart diagram of the proposed calibration method.}
	\label{SystemModel}
\end{figure*}

\section{Mathematical Formulation}

\begin{figure}[!t]
	\centering
	\subfloat[]{\includegraphics[width=4cm]{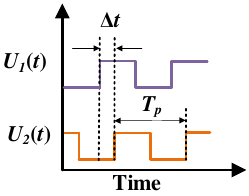}%
		\label{2aa}}
	\hfil
	\subfloat[]{\includegraphics[width=4cm]{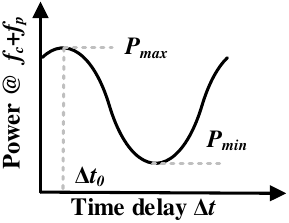}%
		\label{2bb}} 
	\caption{(a) The periodic phase modulation functions of the reference element and the current element, (b) the power of the $+1$st order harmonics varies with the relative time delay.}
	\label{TimeModulationPowerTimeDelay}
\end{figure}
As illustrated in Fig.~\ref{SystemModel}, we consider a uniform linear array under test (AUT) comprising $N$ elements with inter-element spacing d. Each channel contains a phase shifter capable of at least two discrete states ($0^\circ$ and $180^\circ$), with independent FPGA control. Assuming that the far-field calibration source is incident to the AUT at incident angle $\theta$. We designate channel 1 as the reference channel, with its phase modulated by the FPGA between $0^\circ$ and $180^\circ$ at a switching frequency $f_p$ (period $T_p$). The corresponding phase modulation function $U_1(t)$ is defined as

\begin{equation}
	U_{1}(t) = 
	\begin{cases}
		\begin{aligned}
			e^{j0}, &\quad mT_{p} \leq t < \left(m + \frac{1}{2}\right)T_{p} \\
			e^{j\pi}, &\quad \left(m + \frac{1}{2}\right)T_{p} \leq t < (m + 1)T_{p}
		\end{aligned}
	\end{cases},
	\label{U1t}
\end{equation}
where $m\in {{\mathbb{Z}}^{+}}$ represents the cycle index in the periodic phase modulation function.

Similarly, we define the phase modulation function for the $n$th channel under test as $U_n(t)$. While sharing the same modulation period $T_p$ as $U_1(t)$, this modulation function incorporates a programmable time delay $\Delta t$ controlled by the FPGA, as illustrated in Fig.~\ref{TimeModulationPowerTimeDelay}(a). The relationship between these modulation functions can be expressed as

\begin{equation}
	{{U}_{n}}\left( t \right)={{U}_{1}}\left( t-\Delta t \right).
	\label{Unt}
\end{equation}
Then, the synthesized signal received by the array can be expressed as

\begin{equation}
	\begin{aligned}
		S_r(t) &= \left[ U_1(t) A_1 e^{j\varphi_1} + U_n(t) A_n e^{j\varphi_n} e^{jkd(n - 1)\sin\theta} \right] e^{j2\pi f_c t} \\ 
		& + \sum_{i\ne 1,n}^{N} e^{jkd(i - 1)\sin\theta} A_i e^{j\varphi_i} e^{j2\pi f_c t},
	\end{aligned}
	\label{Srt}
\end{equation}
among them, $A_i$ represents the amplitude response of the $i$th element, and $\varphi_i$ represents the phase response of the $i$th element.

When both the reference channel and the channel under test are periodically phase modulated, multiple harmonic components are generated. For frequency-domain analysis, $q$th Fourier coefficient of the reference channel's modulation function ($U_1(t)$) as:

\begin{equation}
	{{\alpha }_{1,q}}=\frac{1}{{{T}_{p}}}\int_{0}^{{{T}_{p}}}{{{U}_{1}}\left( t \right){{e}^{-j2\pi q{{f}_{p}}t}}dt}.
	\label{Alpha1q}
\end{equation}

Since the modulation function of the channel under test is obtained by delaying the reference channel's modulation function, then the corresponding Fourier coefficients exhibit a phase-shift relationship
\begin{equation}
	\begin{aligned}
		{{\alpha }_{n,q}}&=\frac{1}{{{T}_{p}}}\int_{0}^{{{T}_{p}}}{{{U}_{1}}\left( t-\Delta t \right){{e}^{-j2\pi q{{f}_{p}}t}}dt} \\ 
		& ={{\alpha }_{1,q}}{{e}^{-j2\pi q{{f}_{p}}\Delta t}}.
	\end{aligned}
	\label{Alphanq}
\end{equation}
Let ${\Delta t}/{{{T}_{p}}} = \eta ,\left( 0\le \eta <1 \right)$, represents the normalized time delay, such that ${{\alpha }_{n,q}}={{\alpha }_{1,q}}{{e}^{-j2\pi q\eta}}$. This result shows that a time-domain modulation delay $\Delta t$ between the $n$th (test) and the $1$st channels produces a frequency-domain phase shift of $-j2\pi q\eta$ in their respective $q$th harmonic components. The derived relationship demonstrates precise harmonic phase control through relative timing delays between channels. 

Specifically, the phase shift of the $+1$st harmonic exhibits a linear relationship with the relative time delay, expressed as
\begin{equation}
	\Delta {{\varphi}_{+1}}=-2\pi \Delta t / T_p.
	\label{DetlaPhiP1}
\end{equation}
By discretizing the delay step to $\Delta t = T_p/2^{N_{be}}$, where $N_{be}$ is the equivalent phase resolution, the phase of the $+1$st harmonic can be adjusted in increments of $2 \pi /2^{N_{be}} $ radians, matching the precision of an $N_{be}$-bit phase shifter. This equivalence enables high-resolution phase calibration without requiring physical multi-bit phase shifters.

\begin{figure*}[!t]
	\centering
	\subfloat[]{\includegraphics[width=3.9cm]{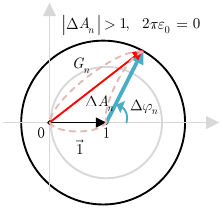}%
		\label{2a}}
	\hfil
	\subfloat[]{\includegraphics[width=4cm]{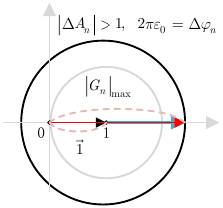}%
		\label{2b}}
	\hfil
	\subfloat[]{\includegraphics[width=4.6cm]{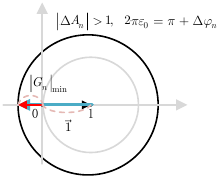}%
		\label{2c}}\\
	\subfloat[]{\includegraphics[width=4cm]{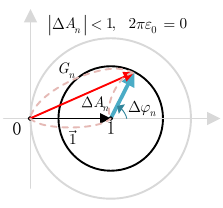}%
		\label{2d}}
	\hfil
	\subfloat[]{\includegraphics[width=4cm]{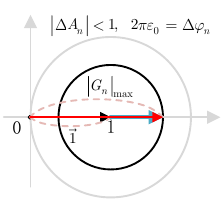}%
		\label{2e}}
	\hfil
	\subfloat[]{\includegraphics[width=4.6cm]{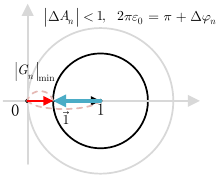}%
		\label{2f}}
	\caption{The schematic of the synthesized vector ${{G}_{n}}$, when the amplitude ratio of the channel under test to the reference channel $\left| \Delta {{A}_{n}} \right|$ is greater than 1 in cases (a-c) or less than 1 in cases (d-f), and the phase difference in the $+1$ harmonic introduced by the time delay and the phase difference between the two channels satisfy (a,d) $2\pi \eta =0$, (b,e) $2\pi \eta =\Delta {{\varphi }_{n}}$, (c,f) $2\pi \eta =\pi +\Delta {{\varphi }_{n}}$ respectively.}
	\label{REHV_AnGreterThan1}
\end{figure*}

Then, in the signal received by the phased array, the $+1$st harmonic can be expressed as
\begin{equation}
	{{H}_{1}}\left( t \right)={{\alpha }_{1,1}}\left( {{A}_{1}}{{e}^{j{{\varphi }_{1}}}}+{{A}_{n}}{{e}^{j{{\varphi }_{n}}}}{{e}^{-j2\pi \eta }} \right){{e}^{j2\pi \left( {{f}_{c}}+{{f}_{p}} \right)t}}.
	\label{H1t}
\end{equation}
The complex response ratio $\Delta A_n = \Delta \gamma _n e^{j\Delta \varphi_n}$ is defined in terms of the amplitude ratio ($\Delta {{\gamma }_{n}}={{{A}_{n}}}/{{{A}_{1}}}$) and phase difference ($\Delta {{\varphi }_{n}}={{\varphi }_{n}}-{{\varphi }_{1}}$) between the $n$th and reference channels, which is what needs to be compensated for in the phased array calibration procedure.
Then, $H_{1}(t)$ can be rewritten as
\begin{equation}
	{{H}_{1}}\left( t \right)={{A}_{1}}{{e}^{j{{\varphi }_{1}}}}{{\alpha }_{1,1}}\left[ 1+\Delta {{\gamma }_{n}}{{e}^{j\left( \Delta {{\varphi }_{n}}-2\pi \eta  \right)}} \right]{{e}^{j2\pi \left( {{f}_{c}}+{{f}_{p}} \right)t}}.
	\label{H1tr}
\end{equation}
The power of the $+1$st harmonic can be expressed as
\begin{equation}
	\begin{aligned}
		 {{P}_{1}(\eta)}&={ {{H}_{1}}\left( t \right)H_{1}^{*}\left( t \right)} \\ 
		& ={{\left| {{A}_{1}}{{\alpha }_{1,1}} \right|}^{2}}\left[ 1+\Delta {{\gamma }_{n}}^{2}+2\Delta {{\gamma }_{n}}\cos \left( \Delta {{\varphi }_{n}}-2\pi \eta  \right) \right]. 
	\end{aligned}
	\label{P1}
\end{equation}

Then, according to (\ref{P1}), it will be find that the power of the $+1$st harmonic varies with the relative modulation time delay between the reference and test channels. Moreover, the power of the $+1$st harmonic varies according to a cosine function with respect to the normalized time delay, as shown in Fig.~\ref{TimeModulationPowerTimeDelay}(b).

\subsection{Phase Calibration}
 According to (\ref{P1}), more importantly, when the power of the $+1$st harmonic is at its maximum, the corresponding time delay and the phase difference ($\Delta {{\varphi }_{n}}$) satisfy the following relationship
\begin{equation}
	\Delta {{\varphi }_{n}}=2\pi \Delta t / T_p.
	\label{DetlaPhin}
\end{equation}
Consequently, this relationship enables direct determination of the phase response difference between the test channel and reference channel using only a periodically modulated 1-bit phase shifter.

It is worth noting that the phase calibration result $\Delta {{\varphi }_{n}}$ depends solely on the time delay $\Delta t$ and is independent of the harmonic coefficients ${{\alpha }_{n,q}}$ and ${{\alpha }_{1,q}}$. When phase shifter errors occur (i.e., the phase switching deviates from the nominal $180^\circ$ shift), the harmonic coefficients are significantly affected. However, since the phase estimation process is independent of these harmonic coefficients, the measurement accuracy of inter-channel phase imbalance remains robust against phase shifter errors.

\subsection{Amplitude Calibration with Ambiguity}
According to (\ref{P1}), the maximum and minimum values of the $+1$st harmonic power are

\begin{equation}
	\left\{ \begin{matrix}
		{{P}_{1\max }}={{\left| {{A}_{1}}{{e}^{j{{\varphi }_{1}}}}{{\alpha }_{1,1}} \right|}^{2}}{{\left( 1+\Delta {{\gamma }_{n}} \right)}^{2}}  \\
		{{P}_{1\min }}={{\left| {{A}_{1}}{{e}^{j{{\varphi }_{1}}}}{{\alpha }_{1,1}} \right|}^{2}}{{\left( 1-\Delta {{\gamma }_{n}} \right)}^{2}}  \\
	\end{matrix} \right..
	\label{P1maxmin}
\end{equation}
where we define the power ratio ${{P}_{r}}={{{P}_{1\max }}}/{{{P}_{1\min }}}\;$. The amplitude ratio $\Delta \gamma_{n}$ can then be determined as

\begin{equation}
	\Delta \gamma_{n} = 
	\begin{cases}
		\dfrac{P_{r}-1}{P_{r}+1}, & \Delta \gamma_{n} \leq 1 \\
		\dfrac{P_{r}+1}{P_{r}-1}, & \Delta \gamma_{n} > 1
	\end{cases}.
	\label{Detlagamman}
\end{equation}
This formulation reveals an inherent ambiguity when calibrating the amplitude inconsistency between channels, a limitation shared with conventional REV methods \cite{mano1982method}.

The amplitude calibration ambiguity inherent in the proposed method can be visualized geometrically through the synthesized vector ${G_n}=1+\Delta \gamma_{n} e^{j(\Delta \varphi_{n}-2\pi \eta)}$, where $\Delta \gamma_{n}$ and $\Delta \varphi_{n}$ represent the amplitude ratio and phase difference between the test channel and reference channel, respectively, and $\eta$ denotes the normalized time delay.

When $2\pi \eta = \varphi_n$ (Fig.~\ref{REHV_AnGreterThan1}(b) and Fig.~\ref{REHV_AnGreterThan1}(e)), the magnitude $\left| {{G}_{n}} \right|$ and the $+1$st harmonic power attain their maxima, irrespective of whether $\left| \Delta {A_n} \right|$ exceeds unity. This property is pivotal for phase calibration, as it ensures a deterministic relationship between the harmonic power peak and the phase difference $\Delta {{\varphi }_{n}}$. Conversely, when $2\pi \eta = \pi + \varphi_n$, (Fig.~\ref{REHV_AnGreterThan1}(c) and Fig.~\ref{REHV_AnGreterThan1}(f)), $\left| {{G}_{n}} \right|$, and the $+1$st harmonic power minimize. For intermediate values of $eta$, the harmonic power varies between these extremes (Fig.~\ref{REHV_AnGreterThan1}(a) and Fig.~\ref{REHV_AnGreterThan1}(d)).

A fundamental limitation arises from this geometric analysis: the harmonic power curve as a function of $eta$ is identical for $\left| \Delta {{A}_{n}} \right| > 1$ and $\left| \Delta {{A}_{n}} \right| < 1$.  Consequently, the method cannot resolve the ambiguity in the amplitude ratio $\Delta {{\gamma }_{n}}$ based solely on the power variation, leading to the calibration ambiguity described by (\ref{Detlagamman}). This challenge mirrors a known limitation in conventional REV methods \cite{mano1982method}, underscoring the need for additional constraints or complementary techniques to uniquely determine $\Delta {{\gamma }_{n}}$.

\subsection{Elimination of the Amplitude Ratio Ambiguity}

To resolve the amplitude calibration ambiguity, we propose a targeted modulation approach where only the $n$th channel of the AUT undergoes periodic phase modulation. The received signal under this configuration is given by

\begin{equation}
	\begin{aligned}
		 {{S}_{rn}}\left( t \right)&=\left[ {{U}_{1}}\left( t \right){{A}_{n}}{{e}^{j{{\varphi }_{n}}}}{{e}^{jkd\left( n-1 \right)\sin \theta }} \right]{{e}^{j2\pi {{f}_{c}}t}} \\ 
		& +\sum\limits_{i\ne n}^{N}{{{e}^{jkd\left( i-1 \right)\sin \theta }}}{{e}^{j2\pi {{f}_{c}}t}}{{A}_{i}}{{e}^{j{{\varphi }_{i}}}}.
	\end{aligned}
	\label{Srn}
\end{equation}
where the first term represents the modulated $n$th channel's contribution, and the second term captures the collective signal from all unmodulated channels. This formulation strategically isolates the channel of interest through its unique modulation signature.

Following the established harmonic analysis framework, we examine the $+1$st harmonic power when solely the $n$th channel is modulated:
\begin{equation}
	{{P}_{1,n}}={{\left| {{\alpha }_{1,1}} \right|}^{2}}A_n^2.
	\label{P1n}
\end{equation}
This reveals a crucial linear relationship: the +1st harmonic power depends exclusively on the squared amplitude response ($A_n^2$) of the modulated channel. This enables direct amplitude characterization through harmonic power measurements, circumventing the ambiguity present in two-channel modulation schemes. The amplitude ratio between the nth channel and reference channel (channel 1) is obtained by normalizing the harmonic powers
\begin{equation}
	\Delta\gamma_n = \sqrt{\frac{P_{1,n}}{P_{1,1}}}, \quad n=2,3,...,N.
	\label{Detagan}
\end{equation}
In addition, according to (\ref{Detagan}), it is available to determine whether $\Delta {{\gamma }_{n}}$ is greater than 1. Then, at the same time, the value of $\Delta {{\gamma }_{n}}$ can be calculated from (\ref{Detlagamman}).

This approach achieves complete calibration of amplitude-phase response inconsistencies across all array elements using only two-phase-state modulation ($0^\circ$ and $180^\circ$) for each phase shifter. Based on the above analysis, only the $+1$st harmonic of the signal received by the AUT needs to be extracted during the entire calibration process. The $+1$th harmonic is generated only by the channel being modulated, and its range of amplitude variation is related solely to the relative amplitude-phase response between the channel under test and the reference channel, independent of the array size. Consequently, this method avoids the drawback that conventional REV methods are more susceptible to noise as the array size increases \cite{10326459}.

\begin{figure}[!t]
	\centering
	\includegraphics[width=6cm]{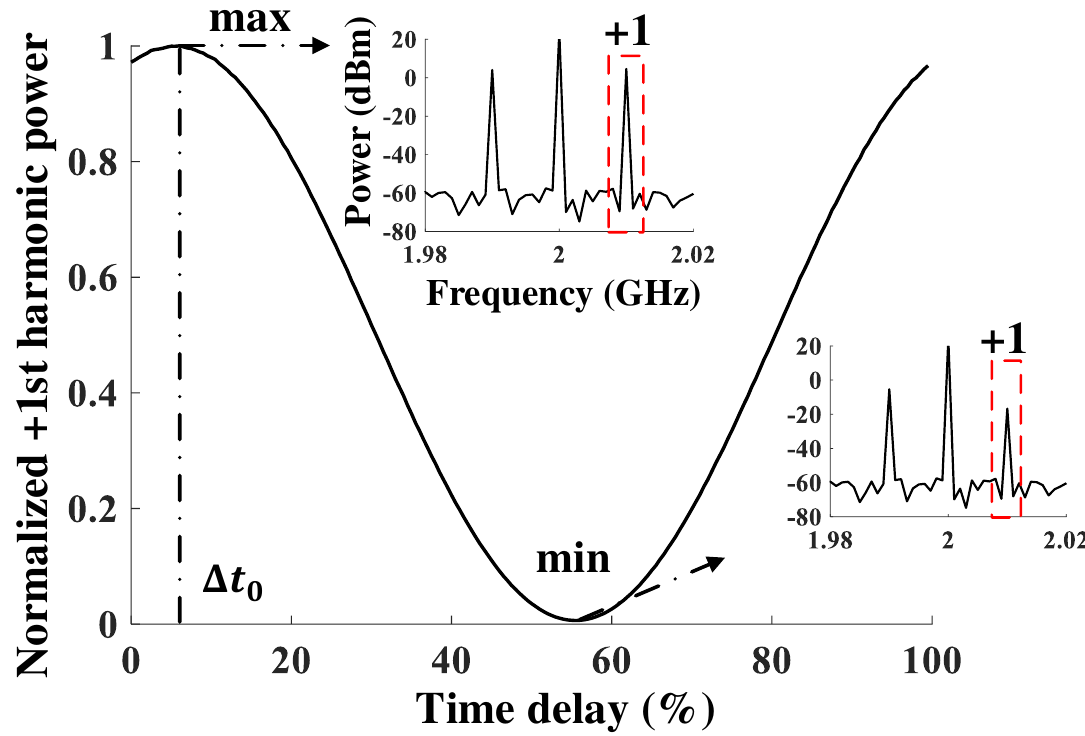}
	\caption{Variation of the power for the $+1$st order harmonics with the relative time delay between the modulation timing of the $2$nd element and the reference element. And the power spectrum of the array-synthesized signal when the $+1$th harmonic power reaches its maximum and minimum values (inset figures).}
	\label{PowerVSDelaySim}
\end{figure}
\section{Numeric Simulations}
This section provides four numerical simulations to demonstrate the workflow of the proposed RHEV calibration method, and evaluates its performance under different SNRs, different AUT sizes and different equivalent phase resolutions.

\subsection{Simulation \Rmnum{1}: workflow of RHEV}

The first numerical simulation illustrates the operational workflow of the RHEV calibration method. As specified in Table \ref{table1}, the simulation employs an eight-element AUT illuminated by a far-field calibration source at 2 GHz with $0^\circ$ incident angle, and an SNR of 20 dB. Each phase shifter undergoes periodic modulation between $0^\circ$ and $180^\circ$ states at 10 MHz. Equivalent phase resolution is 6-bit, which divides each modulation period $T_p$ into $2^6$ intervals, enabling relative time delays between elements to be adjusted in steps of $T_p/2^6$. Element 1 serves as the reference for all measurements.

\begin{table}[htbp]
	\caption{Parameter Settings for Simulation I.}
	\label{table1}
	\centering
	\begin{tabularx}{0.5\textwidth}{>{\centering\arraybackslash}p{0.3\textwidth}>{\centering\arraybackslash}X}
		\toprule
		Parameter & Value \\
		\midrule
		Frequency of the calibration source $f_c$ & 2.0 GHz\\
		SNR & 20 dB\\
		Number of AUT units $N$ & 8 \\
		Phase modulation frequency $f_p$ & 10 MHz \\
		Incident angle $\theta$ & $0^\circ$ \\
		Equivalent phase resolution $N_{be}$ & 6 \\
		\bottomrule
	\end{tabularx}
\end{table}

To determine the phase difference between channel 1 and channel 2, both channels were modulated simultaneously. Figure \ref{PowerVSDelaySim} illustrates the variation in the $+1$st harmonic power with respect to the relative time delay between their phase modulation functions. The harmonic power reaches its maximum at a time delay of 6\%, yielding a phase estimate of $21.6^\circ$, with a deviation of $1.9^\circ$ from the preset value of $19.7^\circ$. 

\begin{figure}[!t]
	\centering
	\includegraphics[width=6cm]{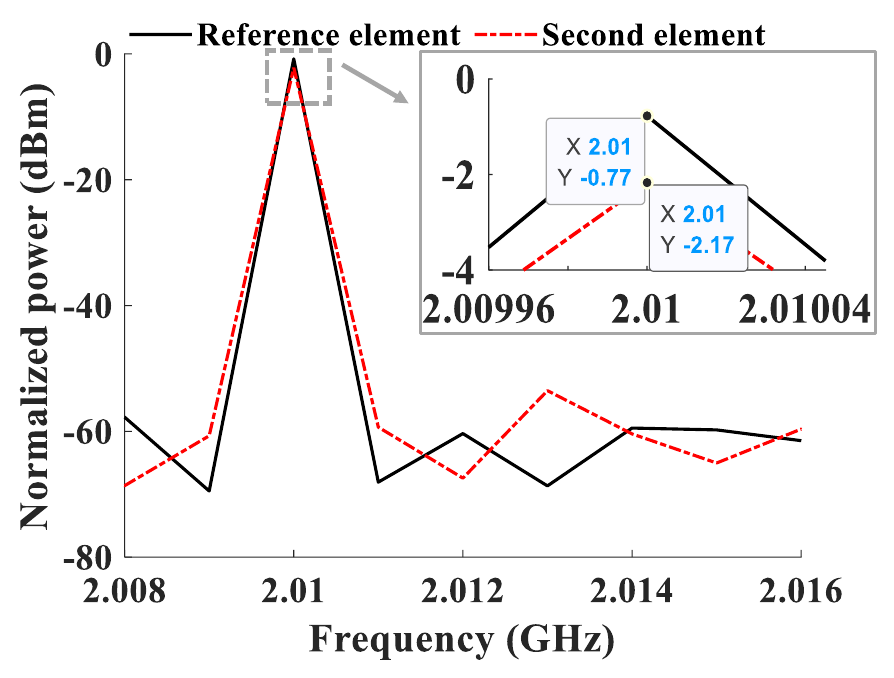}
	\caption{Spectrum of the signal received by the AUT when only one element is periodically phase modulated: the reference element (black solid line), and the second element (red dashed line). The preset value of the amplitude ratio between the second element to that of the reference element is -1.39 dB.}
	\label{REHVSpectrum21}
\end{figure}

For amplitude ratio calibration, we followed the procedure outlined in Section II-C. Each channel was modulated sequentially while measuring the $+1$st harmonic power. The simulated power difference measured -1.397 dBm, matching the preset value of -1.39 dB, as shown in Fig. \ref{REHVSpectrum21}. This result verifies that the amplitude response of channel 2 is indeed lower than that of channel 1. Through (\ref{Detlagamman}), we calculated an amplitude ratio of -1.396 dB, demonstrating the method's effectiveness in resolving amplitude ambiguity.

\begin{figure}[!t]
	\centering
	\includegraphics[width=7cm]{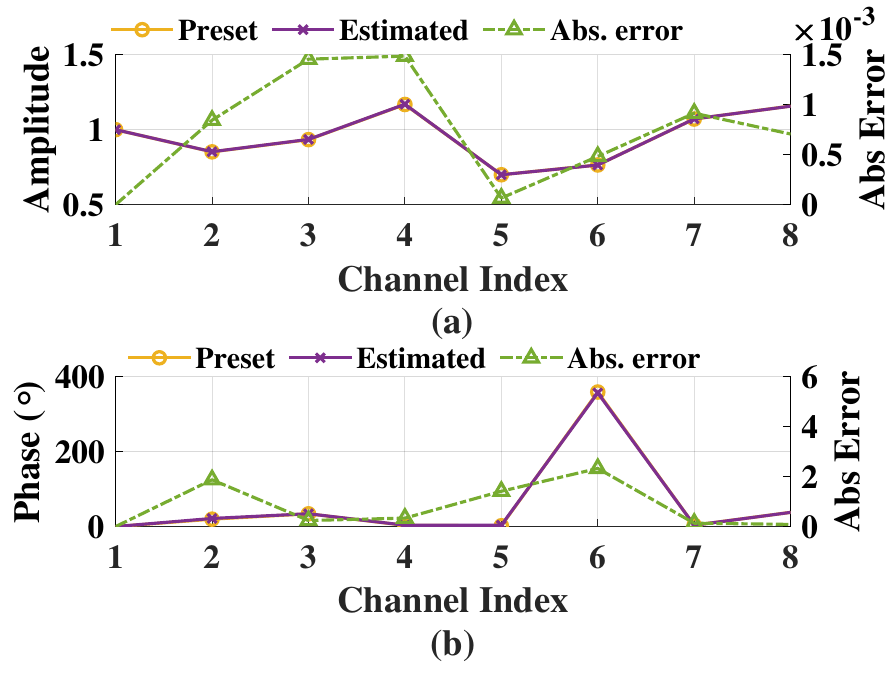}
	\caption{Comparison of (a) the estimated values for amplitude and phase in each channel with the preset values, and (b) the absolute error for the estimated values of amplitude and phase.}
	\label{Sim1PresetVSEsti}
\end{figure}

Finally, Fig.~ \ref{Sim1PresetVSEsti}(a) presents the estimated amplitude ratios and their absolute errors relative to preset values, while Fig. \ref{Sim1PresetVSEsti}(b) shows the corresponding phase difference estimates and errors. For the implemented 6-bit quantization, all observed errors remain within expected tolerances. These results validate the effectiveness of the RHEV method for phased array calibration.

\subsection{Simulation \Rmnum{2}: RMSE with AUT size}

Prior work \cite{10326459} has identified a critical limitation in conventional REV-based calibration methods: the rotating field component's contribution decreases as array size increases, resulting in increased noise sensitivity of the power variation measurements. To quantitatively analyze this effect, we simulate the relationship between SNR and the root mean square error (RMSE) for both amplitude ratio and phase difference estimates across three array configurations (8, 16, and 32 elements). All simulation parameters match those specified in Table~\ref{table1}. The amplitude ratio RMSE is defined as:
\begin{equation}
	\text{RMSE}_{ar}=\frac{1}{N}\sum\limits_{n=1}^{N}{\sqrt{\sum\limits_{m=1}^{M}{{{\left( \frac{{{{\hat{A}}}_{n,m}}}{{{A}_{1,m}}}-\frac{{{A}_{n,m}}}{{{A}_{1,m}}} \right)}^{2}}}}},
\end{equation}
among them, $M$ is the number of Monte Carlo trials, and $m$ indexes individual trials ($1$ to $M$). ${{\hat{A}}_{n,m}}$ is the estimated amplitude response of the $n$th in the $m$th trial. ${A}_{n,m}$ represents the true value of the amplitude response of the $n$th element in the $m$th simulation. $A_{1,m}$ is the reference channel amplitude. $N$ is the number of array elements. Similarly, the RMSE of the estimated phase difference is defined as follows,
\begin{equation}
	\text{RMSE}_{pd}=\frac{1}{N}\sum\limits_{n=1}^{N}{\sqrt{\sum\limits_{m=1}^{M}{{{\left( \frac{{{{\hat{\varphi }}}_{n,m}}}{{{\varphi }_{1,m}}}-\frac{{{\varphi }_{n,m}}}{{{\varphi }_{1,m}}} \right)}^{2}}}}},
\end{equation}
where ${{\hat{\varphi }}_{n,m}}$ and $\varphi _{n,m}$ represent the estimated and true phase differences, respectively.

\begin{figure}[!t]
	\centering
	\subfloat[]{\includegraphics[width=4cm]{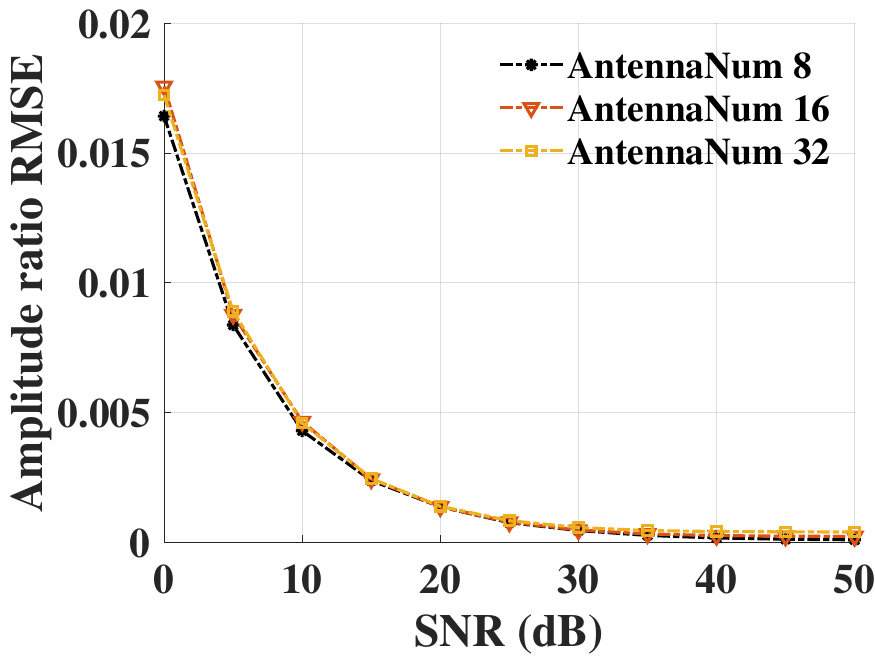}%
		\label{9a}}
	\hfil
	\subfloat[]{\includegraphics[width=4cm]{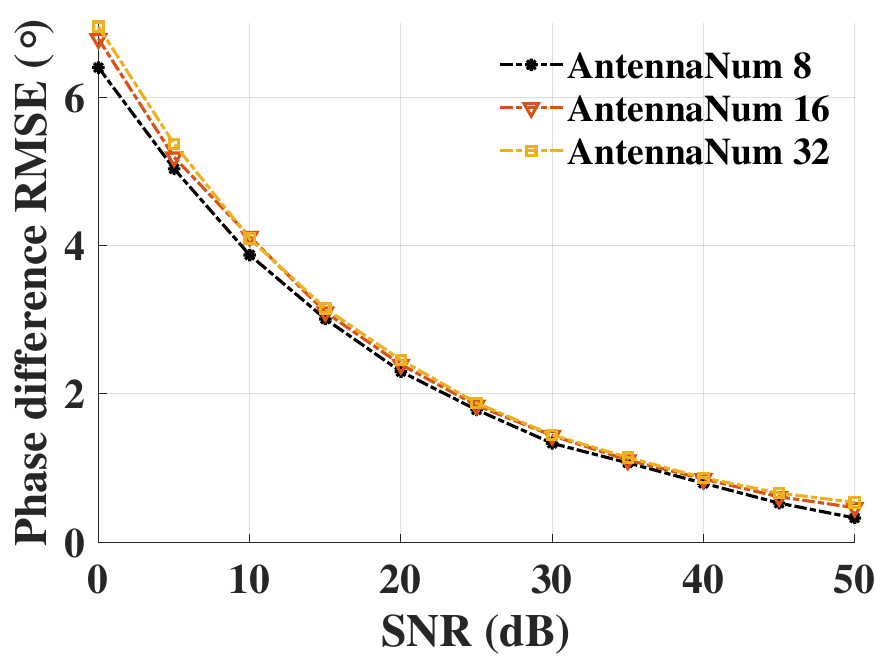}%
		\label{9b}} 
	\caption{Calibration accuracy versus SNR curves under different AUT sizes, (a) amplitude ratio RMSE for RHEV, (b) phase difference RMSE.}
	\label{RMSE_VS_AUTSIZE}
\end{figure}

\begin{figure}[!t]
	\centering
	\subfloat[]{\includegraphics[width=4cm]{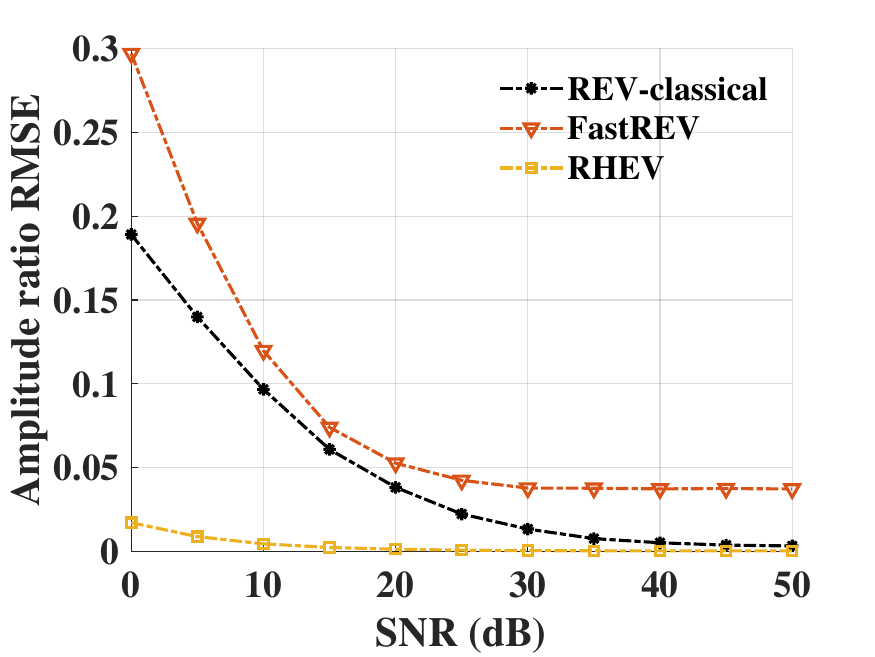}%
		\label{10a1}}
	\hfil
	\subfloat[]{\includegraphics[width=4cm]{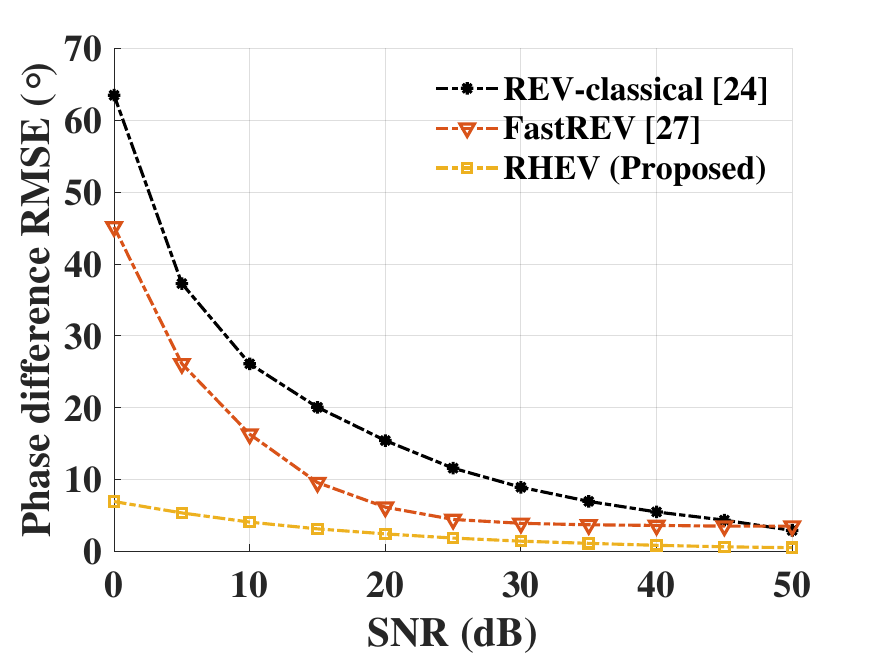}%
		\label{10b1}} 
	\caption{Calibration accuracy (a) amplitude ratio RMSE for RHEV, (b) phase difference RMSE versus SNR curves utilizing different calibration methods when the AUT size is 32. The phase-shift resolution used in the RHEV method and the REV method are both 6-bit.}
	\label{RMSE_VS_METHOD}
\end{figure}

The calibration accuracy versus SNR is analyzed for different array sizes in Fig.~\ref{RMSE_VS_AUTSIZE}, with subplots Fig.~\ref{RMSE_VS_AUTSIZE}(a) and Fig.~\ref{RMSE_VS_AUTSIZE}(b) showing the RMSE of amplitude ratio and phase difference estimates, respectively. Results indicate that both RMSE values decrease monotonically with increasing SNR, while error curves remain nearly identical across all array sizes. This consistency confirms that the RHEV method's calibration accuracy is very little affected by array size, as the method relies exclusively on the $+1$st harmonic for calibration. Unlike the fundamental wave power, which scales with array size, the $+1$st harmonic power variation remains constant.

The performance comparison in Fig.~\ref{RMSE_VS_METHOD} demonstrates the RHEV method's advantages under equivalent 6-bit phase resolution with the same ATU size configuration (32 elements). Subplot Fig.~\ref{RMSE_VS_METHOD}(a) reveals lower amplitude ratio RMSE compared to conventional REV methods, while subplot Fig.~\ref{RMSE_VS_METHOD}(b) shows corresponding improvements in phase difference estimation. 

\subsection{Simulation \Rmnum{3}: RMSE with equivalent phase resolution}

\begin{figure}[!t]
	\centering
	\subfloat[]{\includegraphics[width=4cm]{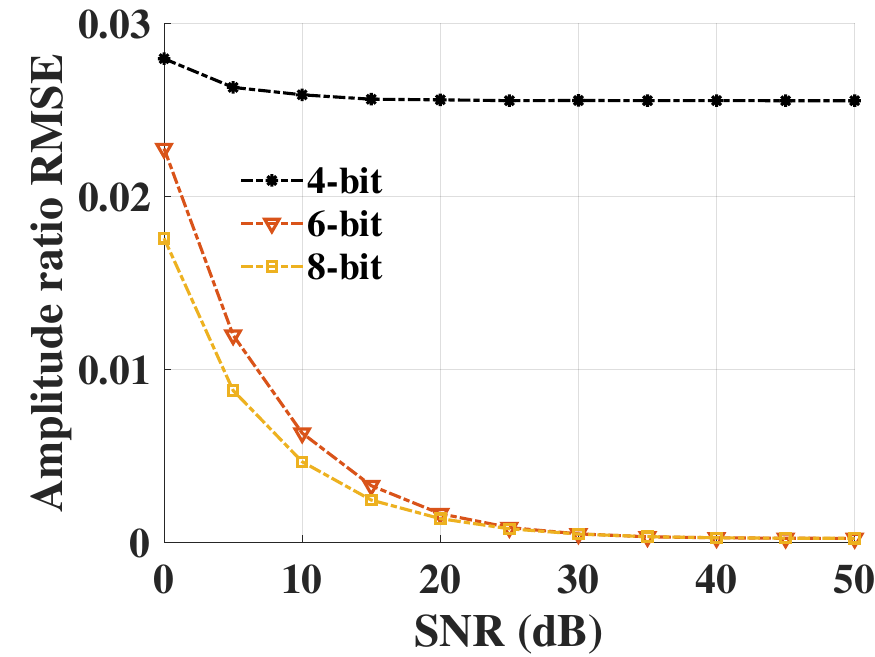}%
		\label{10a}}
	\hfil
	\subfloat[]{\includegraphics[width=4cm]{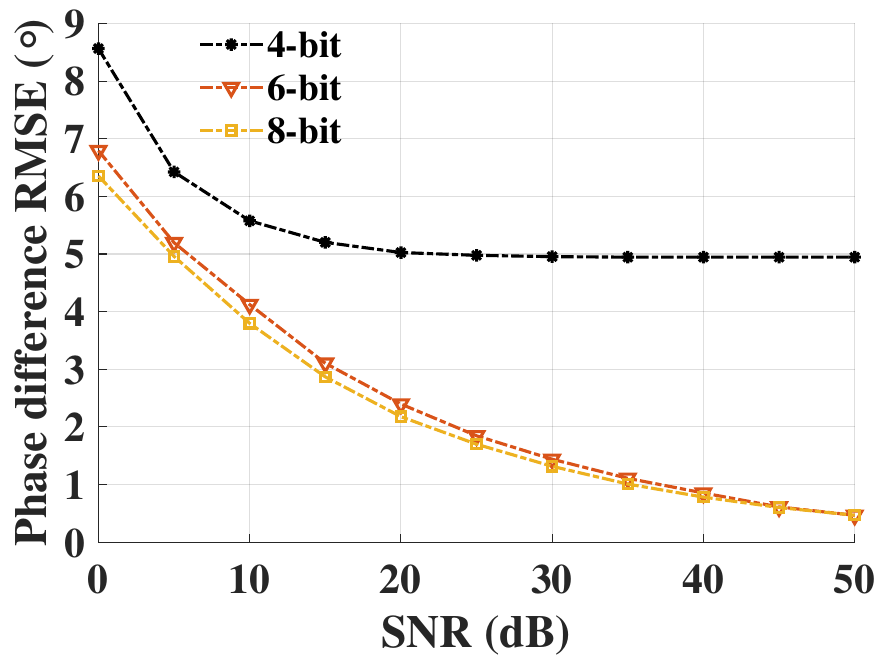}%
		\label{10b}} 
	\caption{Calibration accuracy versus SNR for varying equivalent phase resolutions, (a) amplitude ratio RMSE and (b) phase difference RMSE.}
	\label{RHEV_RMSE_Bits}
\end{figure}
\begin{figure}[t]
	\centering
	\includegraphics[width=6cm]{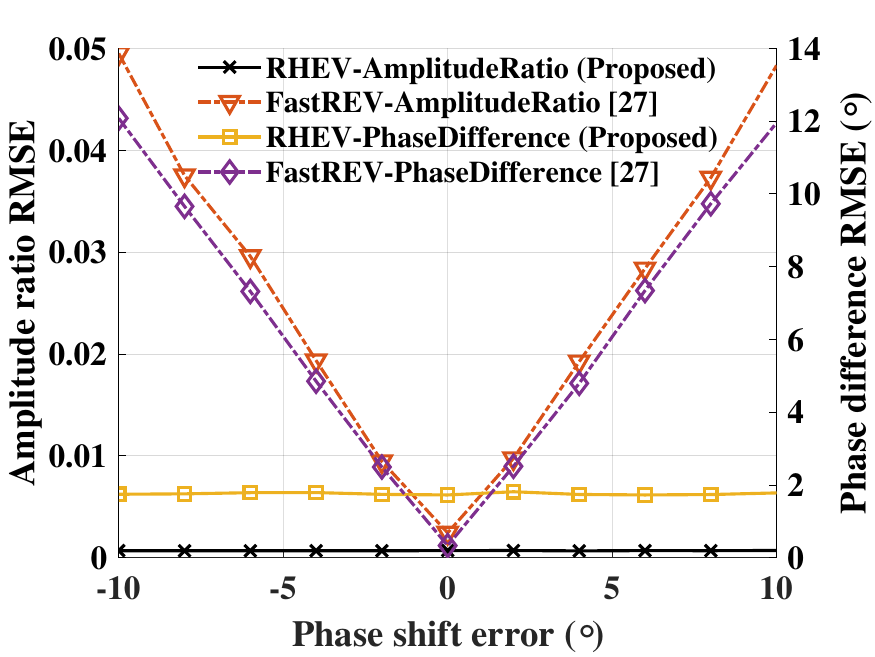}
	\vspace{-2mm}
	\caption{RMSE of (a) amplitude ratio and (b) phase difference estimates with phase shift errors}
	\label{RMSE_VS_PhaseError}
\end{figure}
As shown in Fig.~\ref{RHEV_RMSE_Bits}, demonstrates that calibration accuracy improves with higher equivalent phase resolution (4-bit to 8-bit) but exhibits diminishing returns. At SNR = 15 dB, the phase difference RMSE decreases by approximately 40\% when increasing from 4-bit to 6-bit resolution, while further improvement to 8-bit yields only 8\% additional reduction. For SNR levels above 20 dB, the 6-bit and 8-bit configurations show nearly equivalent performance, with RMSE differences below 5\%, whereas the 4-bit case consistently underperforms by a factor of 2–3 across the entire SNR range. 

Practically, 6-bit resolution offers an optimal tradeoff, delivering 90\% of the phase accuracy and 82\% of the amplitude accuracy achievable with 8-bit at SNR $\ge$ 20 dB, while reducing FPGA resource utilization and improving measurement efficiency. These findings provide quantitative design guidelines.

\subsection{Simulation \Rmnum{4}: RMSE with phase shift error}

Phase shift errors existing in the phase shifter will severely affect the calibration accuracy in the REV \cite{mano1982method}, FastREV \cite{7745942}, and Multiround-grouped-REV methods \cite{10326459}. This is due to the fact that these methods rely on actual accurate phase shifting in their theoretical modeling. The phase shift of the RHEV method, on the other hand, is actually generated by the relative time delay of the modulating signals between the two elements at the $+1$st harmonic, regardless of the phase shift accuracy of the phase shifter itself. Therefore, in the fourth simulation, it is assumed that the 1-bit phase shift error is varied from $-10^\circ$ to $10^\circ$ with $N_{be}=6$, and the rest of the simulation parameters are the same as in Table \ref{table1}. The RMSEs of the amplitude ratio and phase difference caused by different phase shift errors are illustrated in Fig. \ref{RMSE_VS_PhaseError}. It can be noticed that the phase shift error causes only very small fluctuations in the RMSE and does not significantly degrade the calibration accuracy. This phenomenon indicates that the calibration accuracy of the proposed method is insensitive to the phase shift error.

\section{Experiment and Discussion}

To comprehensively validate the proposed method's capability to measure inter-element amplitude-phase imbalances and its OTA calibration performance, we conducted dual experimental investigations: (1) in-channel amplitude-phase response measurement and (2) OTA calibration verification. This approach enables systematic evaluation of both fundamental parameter extraction accuracy and practical calibration efficacy under controlled and real-world conditions, respectively.

\begin{figure}[!t]
	\centering
	\subfloat[]{\includegraphics[width=8cm]{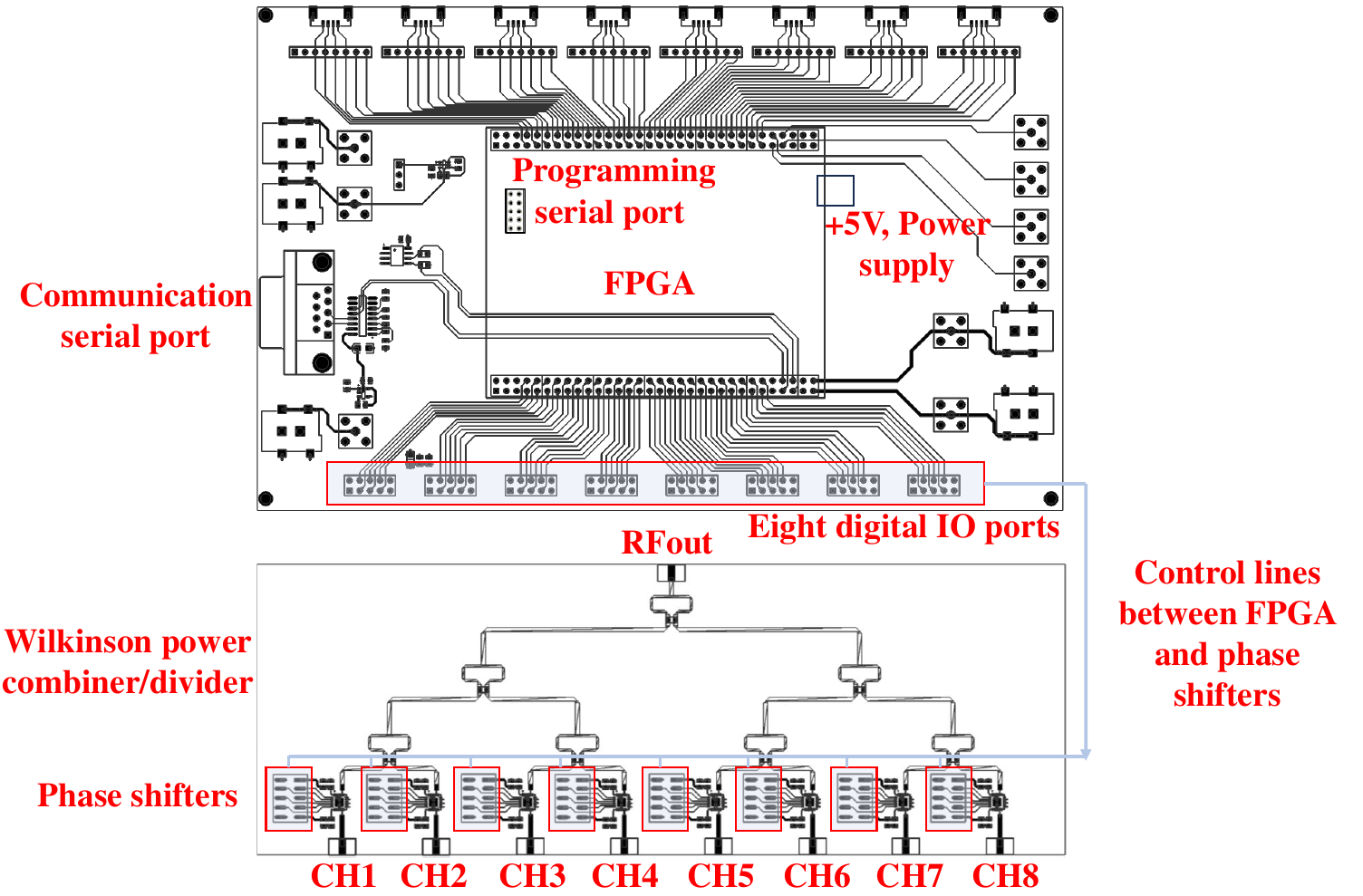}%
		\label{11a}}
	\hfil
	\subfloat[]{\includegraphics[width=6cm]{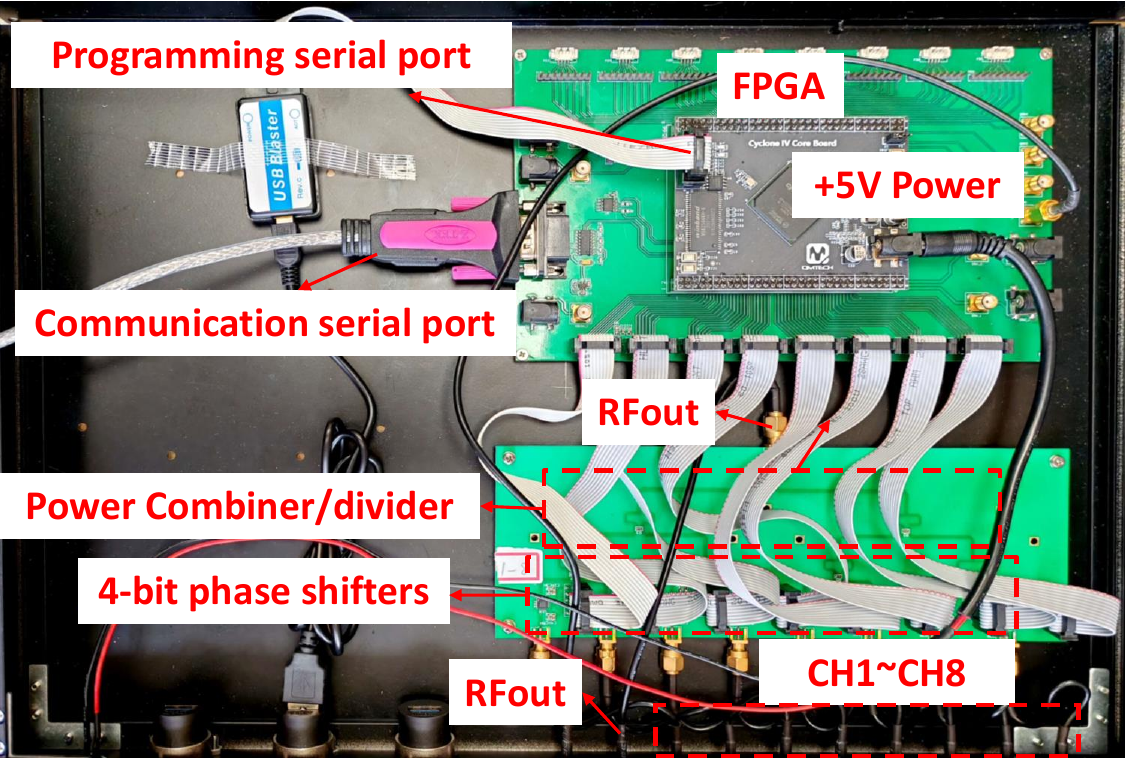}%
		\label{11b}} 
	\caption{(a) Schematic and (b) physical diagram of an eight-channel 4-bit phase shifter network.}
	\label{Exp_4bit_net}
\end{figure}

\begin{figure}[!t]
	\centering
	\subfloat[]{\includegraphics[width=4cm]{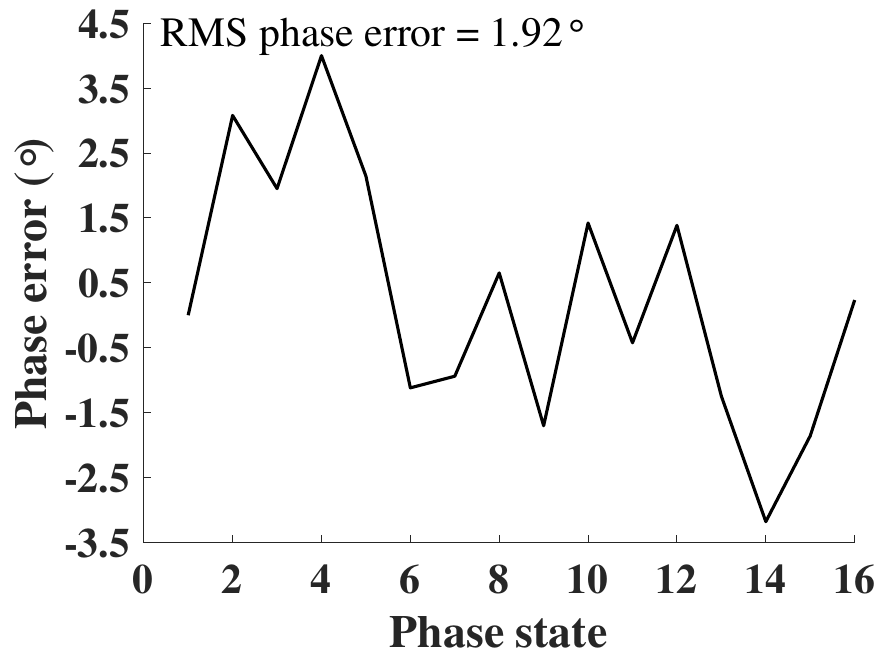}%
		\label{19a}}
	\hfil
	\subfloat[]{\includegraphics[width=4cm]{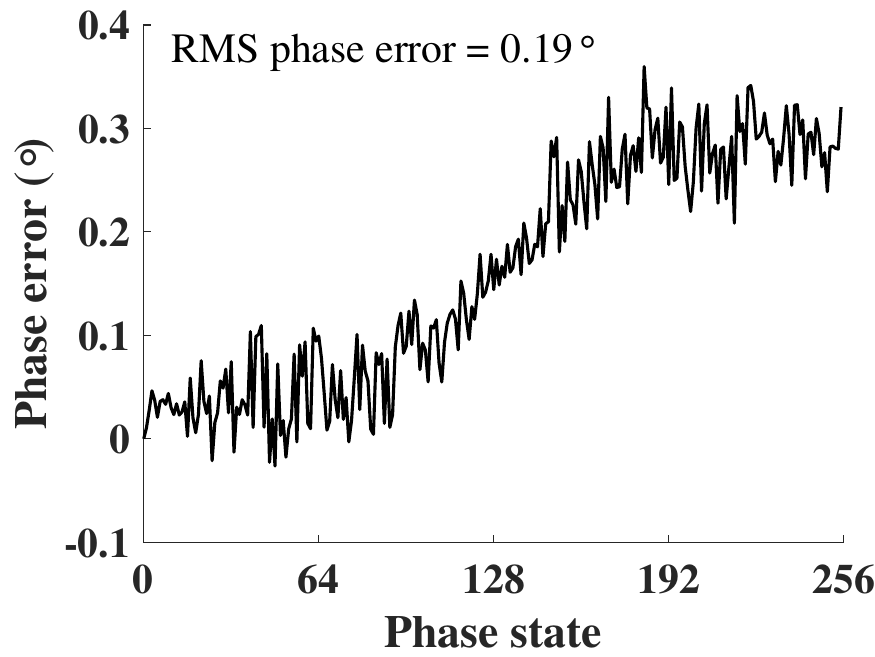}%
		\label{19b}} 
	\caption{Measured phase shift error of, (a) 4-bit physical phase shifter, (b) time-modulated equivalent 8-bit phase shifter.}
	\label{PhaseShifterError}
\end{figure}

\subsection{Calibration in Channel}

\begin{figure}[!t]
	\centering
	\subfloat[]{\includegraphics[width=4cm]{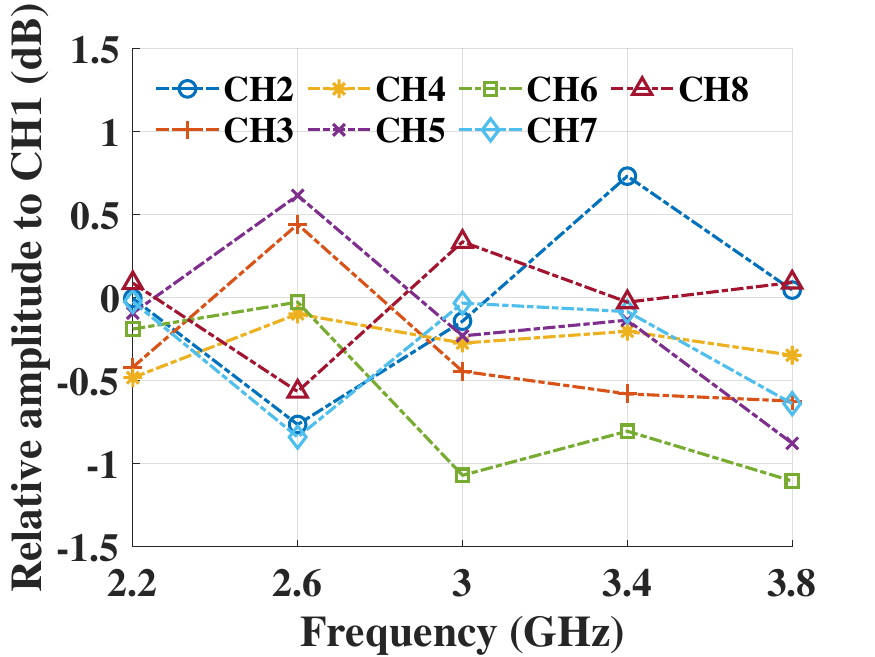}%
		\label{15a}}
	\hfil
	\subfloat[]{\includegraphics[width=4cm]{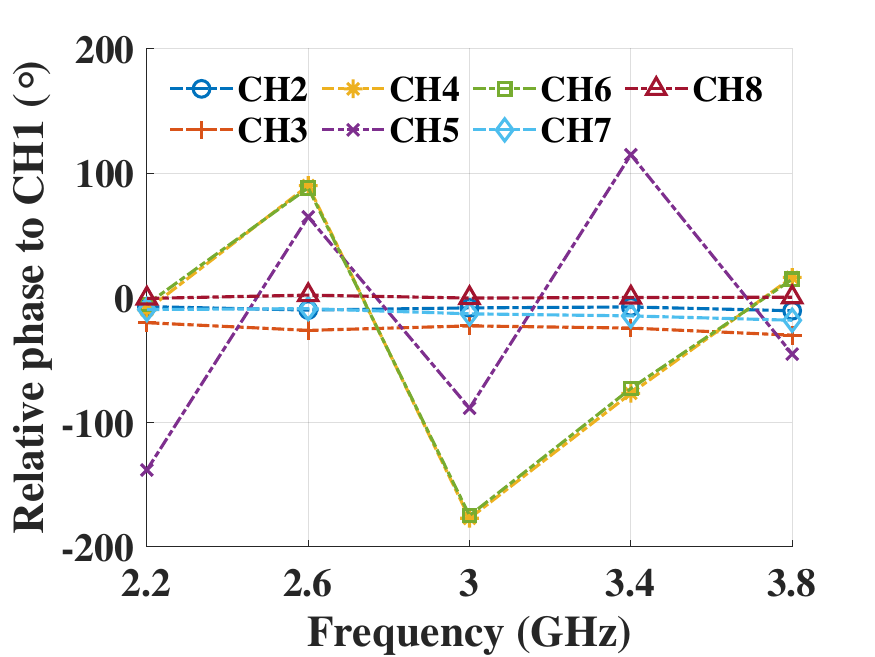}%
		\label{15b}} 
	\caption{(a) Amplitude ratio and (b) phase difference of channels 2 to 8 with respect to channel 1 at different frequencies measured by VNA.}
	\label{InchannelCalibration_S21}
\end{figure}

\begin{figure}[!t]
	\centering
	\subfloat[]{\includegraphics[width=6cm]{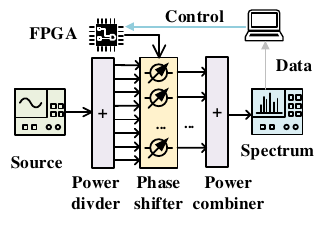}%
		\label{13a}}
	\hfil
	\subfloat[]{\includegraphics[width=6cm]{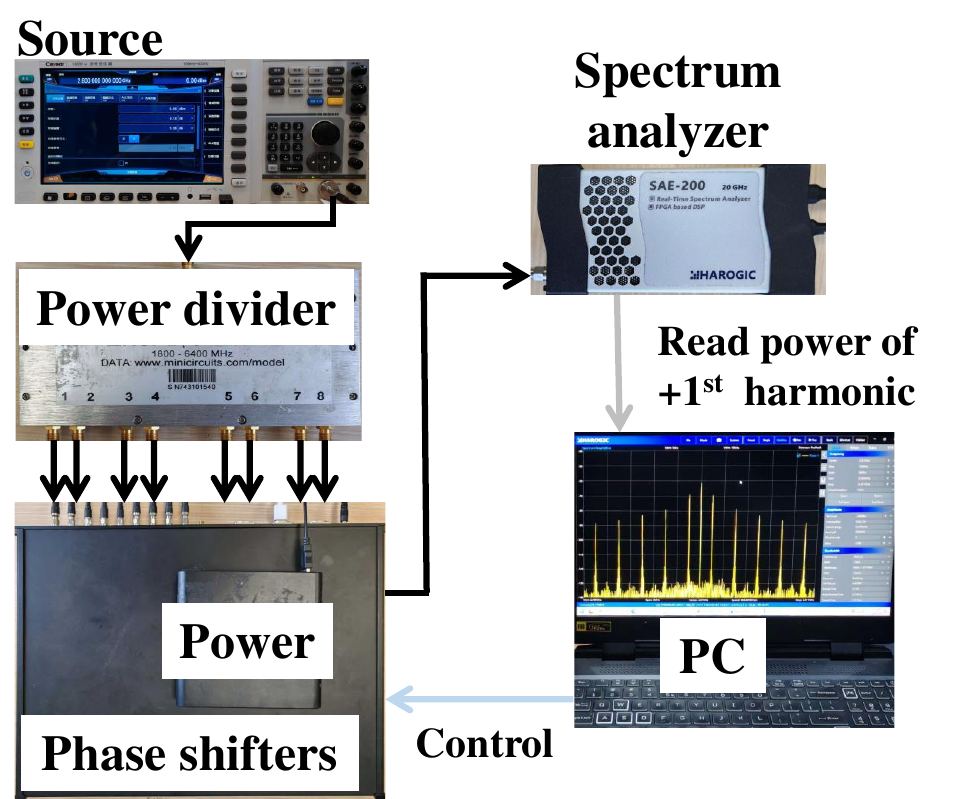}%
		\label{13b}} 
	\caption{(a) Schematic and (b) practical picture for in-channel calibration.}
	\label{InChannel}
\end{figure}
To prevent antenna coupling from affecting the measurements, we performed in-channel tests. The experimental procedure consists of two measurements: equivalent 8-bit phase shift accuracy and amplitude-phase imbalance between channels. The measurement system consists of the following main components:

\textit{(1) Eight-channel 4-bit phase shifter network:} As shown in Fig.~\ref{Exp_4bit_net}, it consists of eight 4-bit digital phase shifters (MAPS-010144, MACOM) operating in the frequency range of 2.2 GHz to 3.8 GHz, an 8-way Wilkinson power combiner (or divider) and an FPGA (EP4CE15F23C8N, Intel). The main functions of the phase shifters are as follows: First, during the measurement stage, two phase states of each phase shifter, $0^\circ$ and $180^\circ$, are selected, and at this time, it is used as a 1-bit phase shifter. Second, during the calibration stage, it is restored to a 4-bit phase shifter. The measurement system employs the Wilkinson combiner in two distinct configurations: (a) as an eight-way power divider for +1st harmonic phase difference measurements, equally distributing the input signal to all eight modulated phase shifters; and (b) as a combiner for amplitude-phase imbalance characterization, recombining the eight phase-shifted/modulated channel outputs into a single composite signal.

The FPGA implementation integrates four critical interfaces to enable full system functionality. A programming serial port handles code configuration, while a separate communication serial port facilitates real-time PC-based control of relative time delays. Eight digital I/O ports provide independent control of each phase shifter channel, complemented by a 5V DC power supply interface. Together, these components support comprehensive system programming, operation, and control.

\textit{(2)RF Signal Source:} It is used to provide a calibration signal source with an operating frequency of 100 kHz~40 GHz (1465F-V, manufactured by Ceyear).

\textit{(3) Power divider:} An one-to-eight power splitter, which can divide the calibration source into eight parts and connect them to the eight phase shifters respectively.

\textit{(4) Portable spectrum analyzer:} The portable spectrum analyzer (HAROGIC SAN200, 9 kHz–20 GHz) measures harmonic power levels in the combined modulated signal, with particular focus on the +1st harmonic component. This instrument interfaces with a personal computer via USB for data acquisition and analysis.

\textit{(5) Personal computer (PC):} The personal computer (PC) controls the FPGA to generate two periodic square waves with programmable time delays while simultaneously acquiring $+1$st harmonic power measurements from the portable spectrum analyzer via USB interface.

\textit{(6) Vector Network Analyzer (VNA):} It is used to test the amplitude-phase response of each channel, that is, the $S_{12}$ parameter, from the output end of the signal source to the input end of the portable spectrum analyzer. Its model is AV3672B-S, manufactured by Ceyear, and the operating frequency range is 10 MHz to 26.5 GHz.
\begin{figure}[t]
	\centering
	\includegraphics[width=6cm]{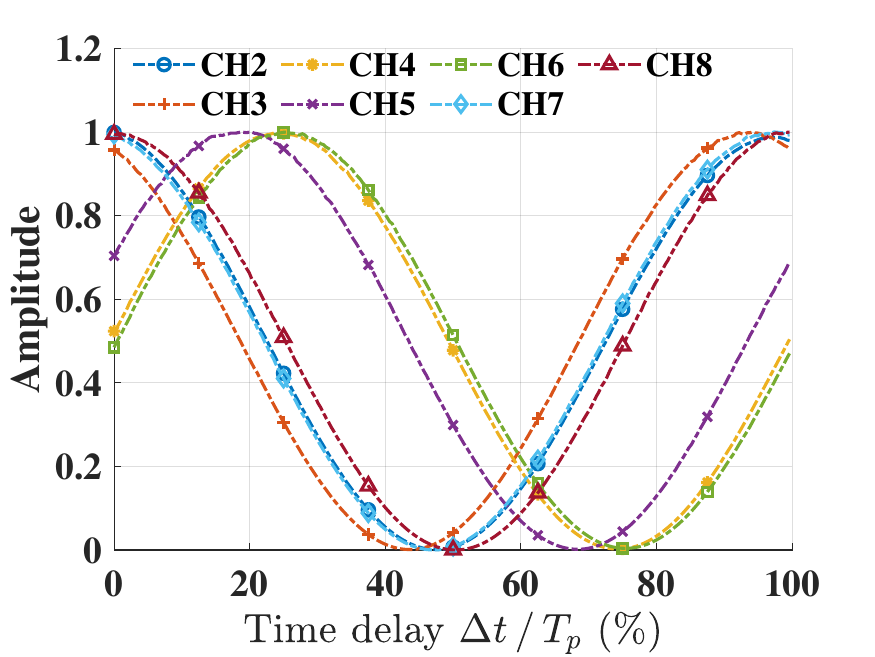}
	\vspace{-2mm}
	\caption{The power of $+1$th harmonic varies with time delay when the calibration source frequency is 2.6 GHz, the equivalent number of bits $N_{be}$ is 8.}
	\label{InchannelCalibration_Fre3G4_PowerP1vsTimeDelay}
\end{figure}

\begin{figure}[!t]
	\centering
	\subfloat[]{\includegraphics[width=4cm]{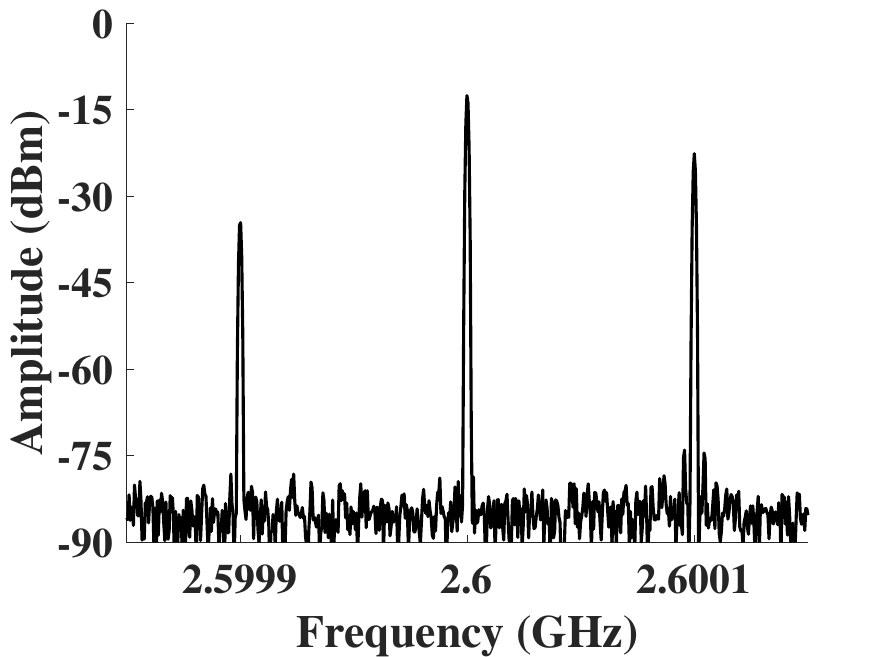}%
		\label{17a}}
	\hfil
	\subfloat[]{\includegraphics[width=4cm]{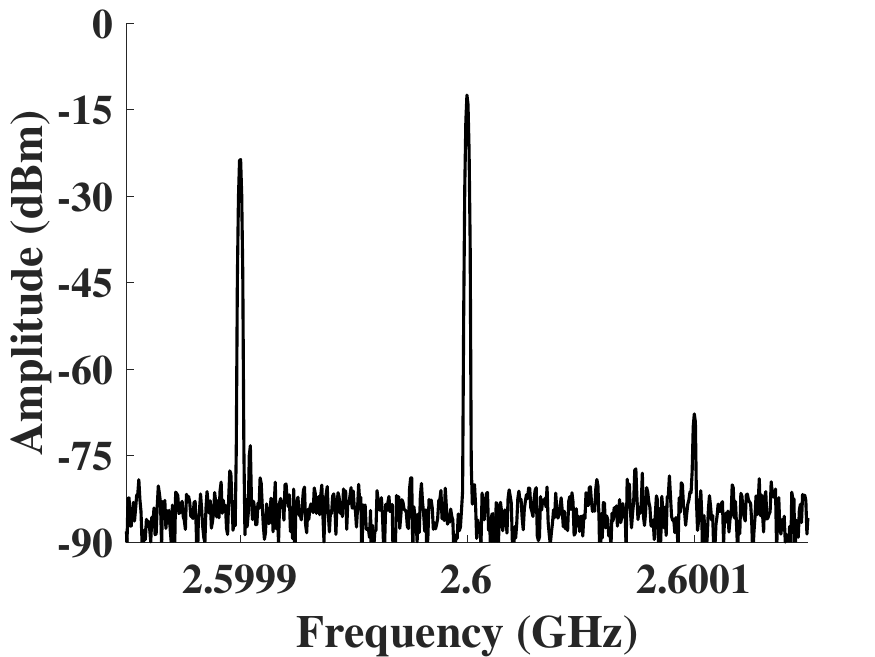}%
		\label{17b}} 
	\caption{Spectrum corresponding to the (a) maximum and (b) minimum of $+1$th harmonic power when the sixth and first channels are modulated periodically at 10 kHz, and the calibration source frequency is 2.6 GHz.}
	\label{Inchannel_spectrum}
\end{figure}

The measurement procedures are as follows: 

\begin{figure}[t]
	\centering
	\includegraphics[width=6cm]{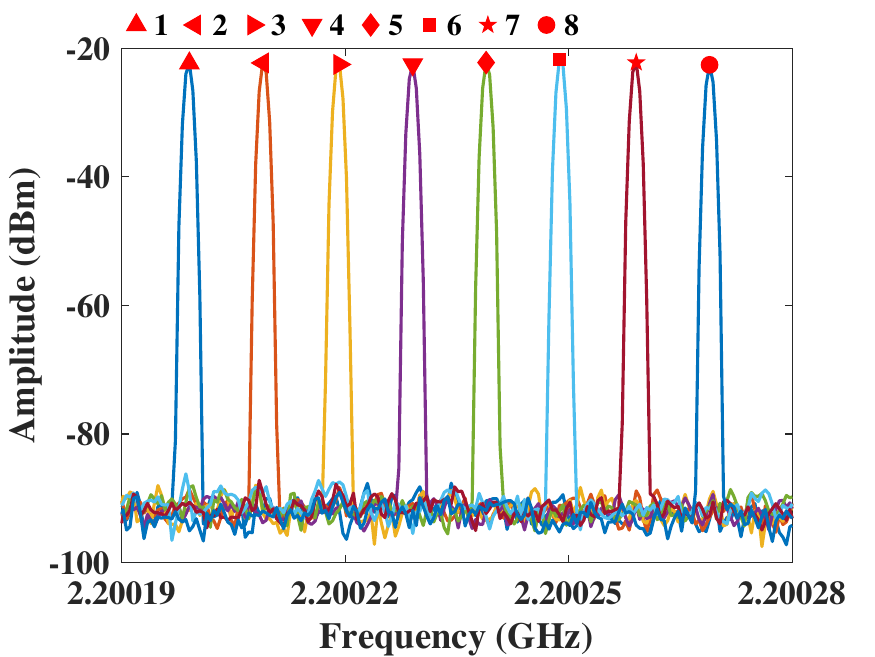}
	\caption{The corresponding $+1$st harmonic power spectra, when sequentially applying periodic phase modulation to elements 1 through 8 at respective frequencies ranging from 200 kHz to 270 kHz in 10 kHz increments.}
	\label{P1harmonic_mea}
\end{figure}
\begin{figure}[!t]
	\centering
	\subfloat[]{\includegraphics[width=4cm]{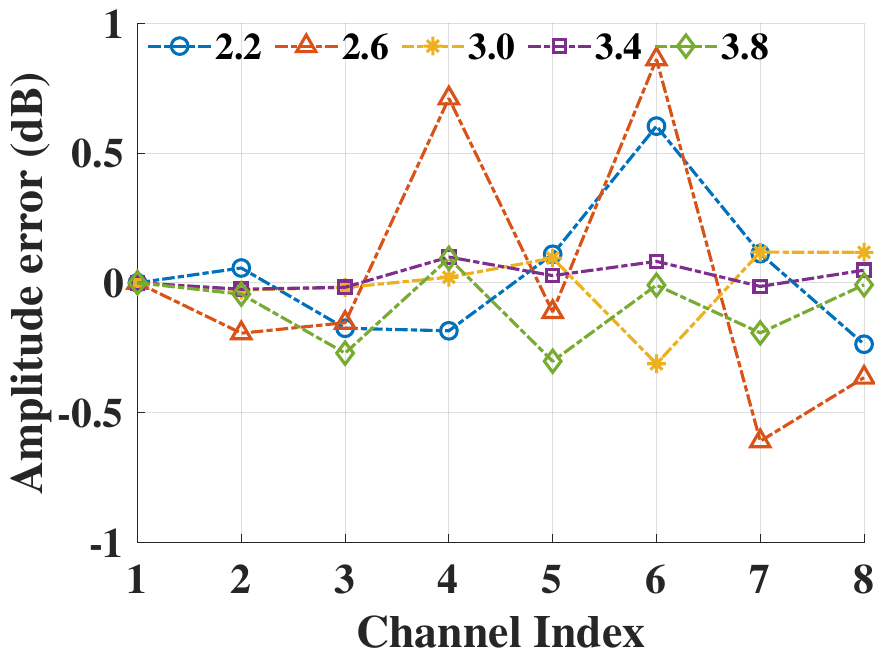}%
		\label{18a}}
	\hfil
	\subfloat[]{\includegraphics[width=4cm]{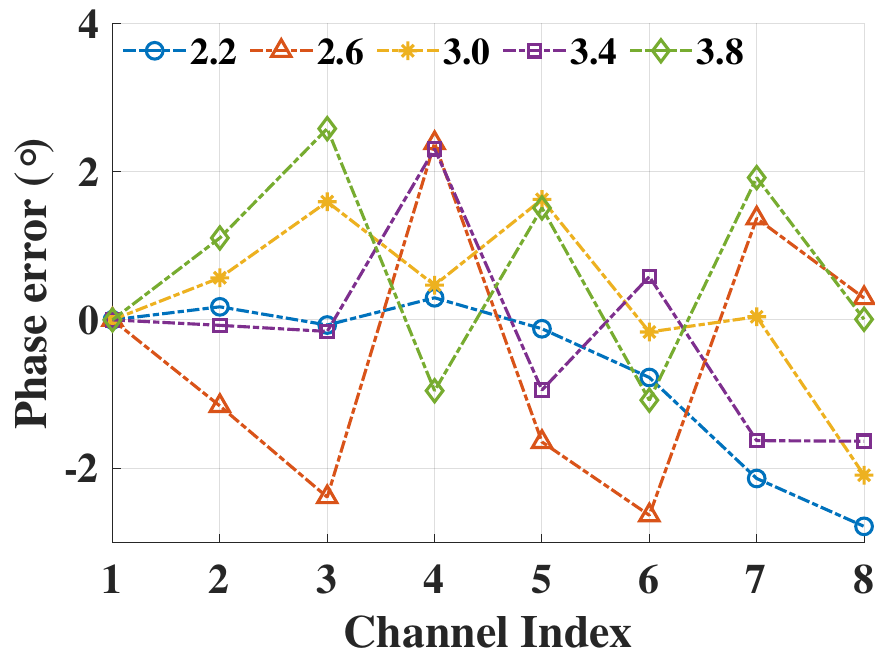}%
		\label{18b}} 
	\caption{The estimation error for the (a) amplitude ratio and (b) phase difference of channels 1 to 8 with respect to channel 1 at different frequencies.}
	\label{InchannelCalibration_Error}
\end{figure}
\textit{(1) Equivalent 8-bit phase shift accuracy measurement:} The Wilkinson combiner was used as a power divider in this experiment. A 2.6 GHz continuous-wave signal at 0 dBm was applied to the input port (RFout as shown in Fig.~\ref{Exp_4bit_net}) of the Wilkinson  divider. The phase shifters in channel 1 and channel 2 were modulated between 0° and 180° at 10 kHz, generating corresponding $+1$st harmonics at 2.60001 GHz in each channel. The FPGA precisely controlled the relative timing delay between the two channels' modulation signals. The phase shifter outputs were connected to the two input ports of VNA. The VNA was configured in receiver mode to measure the phase difference of the 2.60001 GHz ($+1$st harmonic) signals between both ports. The relative modulation delay was incremented in steps of $T_p/2^8$, establishing an equivalent 8-bit phase-shift resolution. Phase differences were recorded across all 256 phase states, with the resulting phase-shift error shown in Fig.~\ref{PhaseShifterError}(b). As a comparison, the 16 static phase shifts of the 4-bit phase shifter in channel 1 were also measured, and the phase shift error results are shown in Fig.~\ref{PhaseShifterError}(a). The time-modulated phase shifting demonstrated superior accuracy, yielding a RMS error of 0.19° at the $+1$st harmonic, representing a tenfold improvement over conventional static phase shifting (RMS error: 1.92°). These results conclusively demonstrate the technique's ability to achieve high-precision harmonic phase control, thereby establishing the theoretical foundation for our proposed REHV-based calibration methodology.

\textit{(2) VNA-based amplitude-phase imbalance measurement:} In this experiment, a VNA was used for measurement, and its frequency range was set within the interval from 2.2 GHz to 3.8 GHz. Port 1 of the VNA is connected to the input terminal of the power splitter via an RF cable. Port 2 is connected to the output of the combiner through an RF cable. In order to measure the amplitude-phase response of the nth channel, the nth output port of the power splitter is connected to the input port of the $n$th ($n = 1, 2, …, 8$) phase shifter via an RF cable. The other output ports of the power divider and the other input ports of the phase shifters are all connected to matching loads. In this way, the amplitude-phase responses of these eight channels can be obtained through the $S_{21}$ measurement function of the VNA. At different frequencies, the amplitude ratios of channels 2 to 8 relative to Channel 1 and the phase differences are shown in Fig.~\ref{InchannelCalibration_S21}(a) and Fig.~\ref{InchannelCalibration_S21}(b) respectively, which are taken as the true values.

\textit{(3) REHV based amplitude-phase imbalance measurement:} Referring to the connection topology shown in  Fig.~\ref{InChannel}(a), firstly, the frequencies of the signal source are set sequentially to 2.2 GHz, 2.6 GHz, 3.0 GHz, 3.4 GHz, and 3.8 GHz, and the output power is set to 0 dBm. Then, it is connected to the input terminal of the power splitter. Subsequently, the output ports of the power splitter are connected to the input terminals of the phase shifters in sequence through RF cables with different lengths. This is aimed at increasing the amplitude-phase imbalance among various channels, the experimental scenario are shown in Fig.~\ref{InChannel}(b). Subsequently, for the phase shifters of the $n$th ($n = 2, 3, \cdots, 8$) channels, phase modulation operations with the same frequency of 10 kHz as that of the first channel are carried out sequentially. During this process, the relative delay between the modulation timing of the $n$th channel and that of the first channel is precisely controlled by FPGA. The minimum steps of the relative delay are quantized to 4 bits, 6 bits, and 8 bits respectively. It is worth noting that at any given moment during the phase calibration, only the $n$th channel and the first channel are periodically phase-modulated at the same frequency, while the phase shifters of the other channels remain static. Finally, connect the output of the combiner to a portable spectrum analyzer, and read the power of the $+1$st harmonic through a PC. Therefore, every time the PC adjusts the relative time delay, it reads the power of the $+1$st harmonic from the spectrum analyzer, thus realizing automated measurement.

During amplitude measurements, each channel undergoes sequential phase modulation at precisely controlled frequencies ranging from 200 kHz to 270 kHz in 10 kHz increments. This periodic modulation scheme is applied to individual channels in a time-division sequence. The $+1$st harmonic power spectra generated by each element are presented in Fig.~\ref{P1harmonic_mea}. The ambiguity in amplitude ratio can be resolved by analyzing the relative power levels of the $+1$st harmonics generated from each modulated element.

\textit{(3) Measurement result:} Suppose the frequency of the calibration source is set to 2.6 GHz. When the sixth channel and the first channel are periodically modulated at the same time, the spectrums corresponding to the time when the $+1$st harmonic power reaches the maximum and minimum values, are shown in Fig.~\ref{Inchannel_spectrum}(a) and Fig.~\ref{Inchannel_spectrum}(b), respectively. The changes of the $+1$st harmonic's power caused by the relative time delays are shown in Fig.~\ref{InchannelCalibration_Fre3G4_PowerP1vsTimeDelay}. It can be clearly found that there is a typical sinusoidal variation pattern between the power of the $+1$st harmonic and the relative time delay. According to Fig.~\ref{InchannelCalibration_S21}(b), when the frequency is 2.6 GHz, the phase of channel 4 is about $1.8^\circ$ forward than that of channel 6. Thean, turning our attention to Fig.~\ref{InchannelCalibration_Fre3G4_PowerP1vsTimeDelay}, the $+1$st harmonic power of channel 4 peaks about 0.5\% of a modulation period earlier than that of channel 6. This indicates that the phase difference between channel 4 and channel 6 is about $0.5\%\times360^\circ = 1.8^\circ$, which is consistent with the VNA measurement. Intuitively, the two curves overlap almost exactly, which also indicates that the amplitude-phase response of these two channels is approximately equal. In addition, within the operating frequency range of the phase shifter, when the frequencies are 2.2 GHz, 3.0 GHz, 3.4 GHz, and 3.8 GHz, the measurement errors of the amplitude-phase responses between channels are shown in Fig.~\ref{InchannelCalibration_Error}(a) and Fig.~\ref{InchannelCalibration_Error}(b) respectively. The amplitude error is between $\pm0.9$ dB, and phase error is between $\pm2.8^\circ$. As shown in Table~\ref{table2}, decreasing the number of equivalent quantization bits to 6-bit and 4-bit, respectively, the phase error and amplitude error increase, but the errors corresponding to 6-bit and 8-bit are closer, and the error corresponding to 4-bit is larger, which is consistent with the simulation results.

\begin{table}[htbp]
	\centering
	\caption{Comparison of Estimation Errors between Different Calibration Methods}
	\label{table2}
	\begin{tabular}{>{\centering\arraybackslash}p{1.5cm} >{\centering\arraybackslash}p{1.5cm} >{\centering\arraybackslash}p{1.5cm} >{\centering\arraybackslash}p{1.5cm}} 
		\toprule
		Method & Amplitude ratio error (dB) & Phase difference error (dB) & Amplitude-only\\ 
		\midrule
		\cite{mano1982method}  & $\pm$1.3 dB &  $\pm7.7^\circ$ & \checkmark \\ 
		\cite{7745942}  & $\pm$1.4 dB &  $\pm9.6^\circ$ & \checkmark\\
		\cite{10326459}  & $\pm$1.4 dB &  $\pm7.3^\circ$ & \checkmark\\
		\cite{6868196} & $\pm$2 dB &  $\pm5.8^\circ$ & \ding{55} \\
		\cite{10310232} & $\pm$1.2 dB &  $\pm6.2^\circ$ & \ding{55}\\
		\textbf{REHV/8-bit} & $\pm$0.9 dB & $\pm2.8^\circ$  & \checkmark\\
		\textbf{REHV/6-bit} & $\pm$1.1 dB & $\pm2.9^\circ$  & \checkmark\\
		\textbf{REHV/4-bit} & $\pm$1.2 dB & $\pm4.6^\circ$  & \checkmark\\
		\bottomrule
	\end{tabular}
\end{table}

Compared to other amplitude-only calibration methods, the proposed approach achieves minimum measurement errors in both amplitude ratio and phase difference when the equivalent bit resolution is 6. Even at 4-bit equivalence, it maintains superior accuracy over conventional REV methods. However, similar to traditional REV techniques, this method requires a relatively high number of measurements. Addressing this limitation by optimizing measurement efficiency will be a key focus of future research.

\begin{figure}[!t]
	\centering
	\includegraphics[width=8cm]{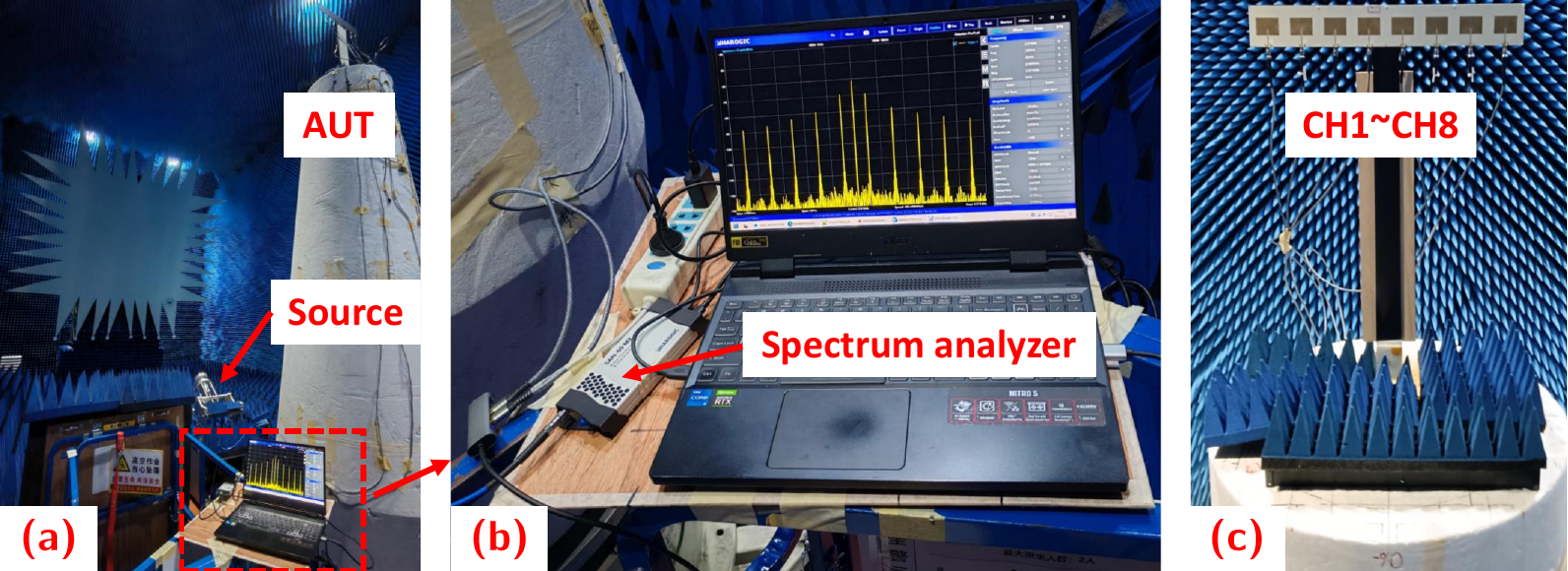}
	\caption{The experimental scenarios for OTA calibration in a compact range microwave anechoic chamber. (a) The overall experimental scenario, (b) The receiving end, (c) The AUT and the eight-channel 4-bit digital phase shifter.}
	\label{Expall}
\end{figure}
\begin{figure}[!t]
	\centering
	\includegraphics[width=6cm]{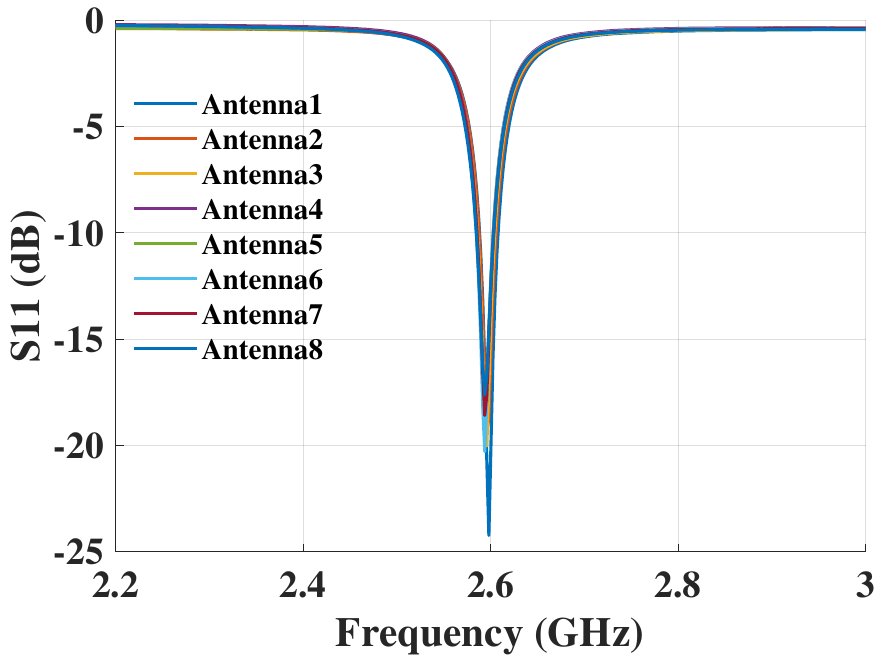}
	\caption{The return loss ($\text{S}_{\text{11}}$ parameter) of the employed eight patch antennas.}
	\label{AntennaS21}
\end{figure}

\subsection{ Calibration OTA}

\begin{figure}[!t]
	\centering
	\includegraphics[width=7cm]{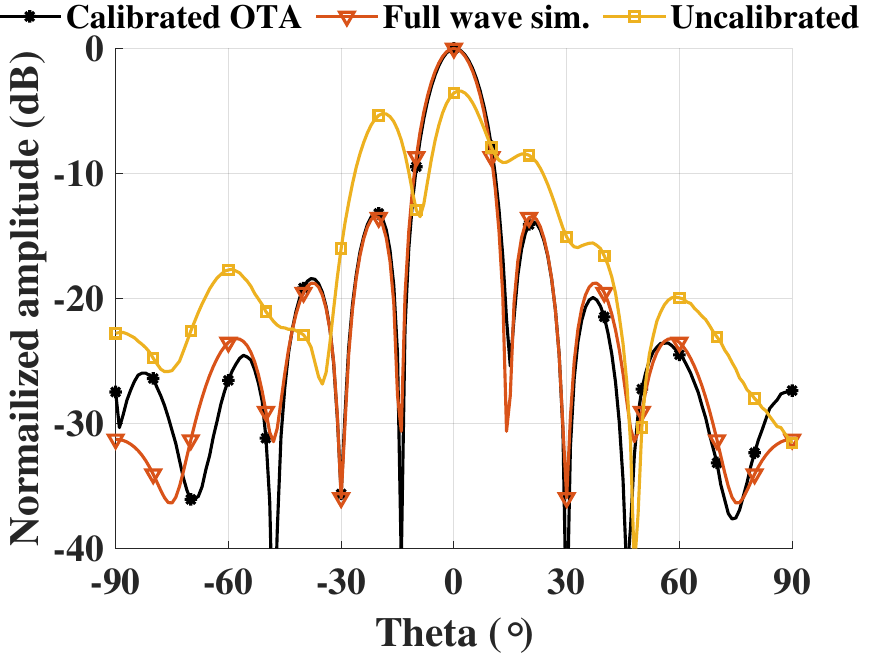}
	\caption{The power patterns of the array when the beam is pointed at $0^\circ$ before and after calibration.}
	\label{CalibrationOTAScenarios}
\end{figure}

\begin{figure}[!t]
	\centering
	\includegraphics[width=7cm]{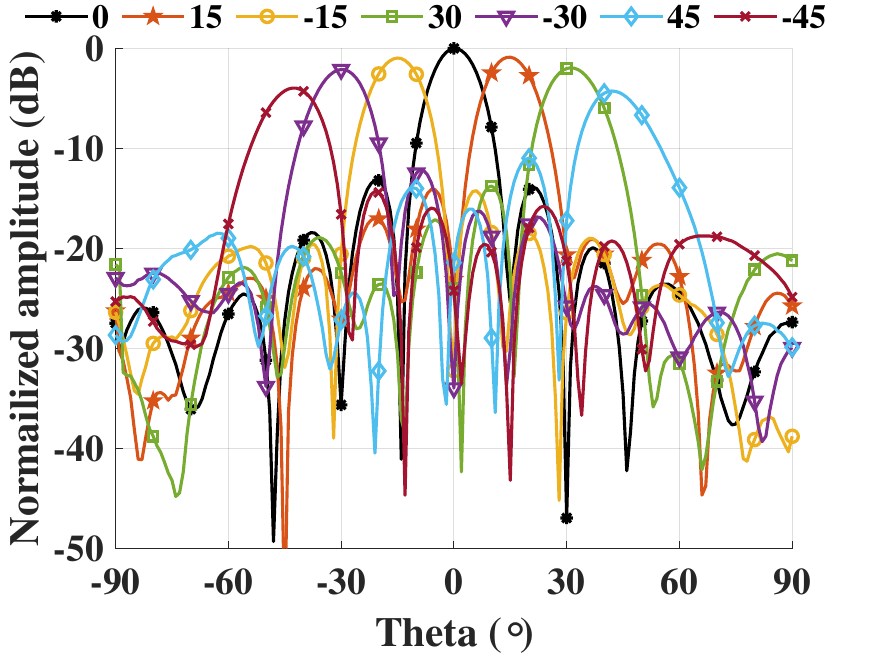}
	\caption{The power patterns of the array when the beam is pointed at $0^\circ$ before and after calibration.}
	\label{PatternBeamsAfterCalibration}
\end{figure}
To validate the effectiveness of the REHV method for OTA calibration, we fabricated an 8-element uniform linear array operating at 2.6 GHz, which is shown in as shown in Fig.~\ref{Expall}(c). The return loss (S11 parameter) characteristics of the eight patch antennas are shown in Fig.~\ref{AntennaS21}. Different from in-channel calibration, an antenna array is used to replace the eight-way power divider as the input of the phase shifter. The eight antennas are sequentially connected to the input ends of the eight-way phase shifter through RF cables with different lengths. As shown in Fig.~\ref{Expall}(a), in the compact range anechoic chamber, the AUT is placed on the turntable in the quiet zone. As shown in Fig.~\ref{Expall}(b), connect the output of the combiner to a portable spectrum analyzer. The PC is connected to the portable spectrum analyzer and the FPGA respectively via USB. The calibration source has a power of 10 dBm and a frequency of 2.6 GHz. The remaining test steps are the same as those for in-channel calibration. 

When the calibration signal is incident vertically on the AUT, the corresponding normalized power patterns before and after calibration are obtained, and their comparison is shown in Fig.~\ref{CalibrationOTAScenarios}. Through analysis, it can be known that the main beam of the pattern after calibration can accurately point to $0^\circ$. Moreover, the antenna gain has been increased by 3.4 dB, and the sidelobe level has been reduced by 11.4 dB compared with that before calibration. After the calibration is completed, the amplitude-phase response of each element are taken as the excitation parameters for subsequent full wave electromagnetic simulation by CST MWS. By comparing the normalized power pattern obtained from the simulation with the measured pattern after calibration (as shown in Fig.~\ref{CalibrationOTAScenarios}), the two highly coincide in key characteristics such as the main beam direction, sidelobes, and null distribution.

In addition, to further verify the calibration effect, the beam scanning capability of the antenna array after calibration within the range from $-45^\circ$ to $45^\circ$ is measured, and the results are shown in Fig.~\ref{PatternBeamsAfterCalibration}. The experimental data and simulation results confirm each other, fully demonstrating that the RHEV method has remarkable effectiveness and practical value in the field of OTA calibration.

\section{Conclusion}
The proposed RHEV-based calibration method presents an innovative solution for phased array calibration that overcomes key limitations of conventional calibration methods with amplitude-only measurements. The core idea of the proposed method is to realize high-precision phase control at the $
+1$st harmonic by controlling the relative time delay of the modulation timing between the two cyclically modulated phase shifters. Since the phase shift is generated by the delay rather than the physical phase shifters, the performance of RHEV-based calibration is not susceptible to phase shifter errors. Furthermore, by exclusively utilizing the $+1$st harmonic for calibration while avoiding the fundamental frequency component, the proposed method maintains consistent calibration performance regardless of array size. The demonstrated ability to achieve an equivalent 8-bit phase shift with 0.19° RMSE at the $+1$st harmonic provides the theoretical foundation for this method. Experimental results confirm the method's effectiveness, showing in-channel amplitude and phase imbalances below 0.9 dB and 2.8° respectively. Furthermore, successful OTA calibration of a 2.6 GHz phased array validates the practical implementation of this technique. Future work will focus on optimizing the calibration efficiency of the proposed method.

\ifCLASSOPTIONcaptionsoff
  \newpage
\fi



\bibliographystyle{IEEEtran}
\bibliography{Coherent_Calibration}

\end{document}